\documentclass{article}                     
\usepackage{graphicx}
\usepackage{a4wide}             
\usepackage{color}              
\usepackage{amsmath}            
\usepackage{amssymb}            
\usepackage[colorlinks=true,linkcolor=azulondos,citecolor=azulondos,urlcolor=azulondos]{hyperref}
\usepackage{breakcites}

\definecolor{verdoso}{rgb}{0.1,0.4,0.1}
\definecolor{azulondos}{rgb}{0,.33,.61}

\newcommand{\ba}{\mathbf{a}}
\newcommand{\bj}{\mathbf{j}}
\newcommand{\bl}{\mathbf{l}}
\newcommand{\br}{\mathbf{r}}
\newcommand{\bu}{\mathbf{u}}
\newcommand{\bx}{\mathbf{x}}

\newcommand{\bz}{\mathbf{z}}

\newcommand{\bA}{\mathbf{A}}
\newcommand{\bB}{\mathbf{B}}
\newcommand{\bI}{\mathbf{I}}

\newcommand{\bzero}{\mathbf{0}}

\newcommand{\bEpsilon}{\boldsymbol{\varepsilon}}
\newcommand{\bgamma}{\boldsymbol{\gamma}}
\newcommand{\bGamma}{\boldsymbol{\Gamma}}
\newcommand{\bPhi}{\boldsymbol{\Phi}}
\newcommand{\bPsi}{\boldsymbol{\Psi}}
\newcommand{\bSigma}{\boldsymbol{\Sigma}}

\newcommand{\E}{\text{E}}
\newcommand{\Cov}{\text{Cov}}
\newcommand{\Var}{\text{Var}}

\begin{document}

\title{A tutorial on reproducing a predefined autocovariance function through AR models: Application to stationary homogeneous isotropic turbulence}

\maketitle

\noindent \large{Cristobal Gallego-Castillo$^{1,*}$,  Alvaro Cuerva-Tejero$^1$, Mohanad Elagamy$^1$, }

\noindent \large{Oscar Lopez-Garcia$^1$, Sergio Avila-Sanchez$^1$ } 

\

\noindent \small{$^1$DAVE (ETSIAE), Universidad Politécnica de Madrid. Pza. Cardenal Cisneros, 3, 28040, Madrid}

\vspace{0.2cm}

\noindent \small{$^*$Corresponding author: Cristobal Gallego-Castillo.  

\noindent $~$ Email: \href{mailto:cristobaljose.gallego@upm.es}{\texttt{cristobaljose.gallego@upm.es}}. 

\noindent ~ ORCID:\href{https://orcid.org/0000-0002-8249-5179}{0000-0002-8249-5179} }


\begin{abstract}

Sequential methods for synthetic realisation of random processes have a number of advantages compared with spectral methods. In this article, the determination of optimal autoregressive (AR) models for reproducing a predefined target autocovariance function of a random process is addressed. To this end, a novel formulation of the problem is developed. This formulation is linear and generalises the well-known Yule-Walker (Y-W) equations and a recent approach based on restricted AR models (Krenk-M{\o}ller approach, K-M). Two main features characterise the introduced formulation: (i) flexibility in the choice for the autocovariance equations employed in the model determination, and (ii) flexibility in the definition of the AR model scheme. Both features were exploited by a genetic algorithm to obtain optimal AR models for the particular case of synthetic generation of homogeneous stationary isotropic turbulence time series. The obtained models improved those obtained with the Y-W and K-M approaches for the same model parsimony in terms of the global fitting of the target autocovariance function. Implications for the reproduced spectra are also discussed. The formulation for the multivariate case is also presented, highlighting the causes behind some computational bottlenecks.

\


\end{abstract}


\section{Introduction} \label{sec_10_intro}

The synthetic realisation of a random process (or, simply, the numerical generation of a random process) refers to the computational generation of time/space series that simulate the random behaviour of a dynamical system accurately. In this context, accurately means that the obtained time/space series reproduce sufficiently well a number of statistical features defined beforehand, like time/space covariance and cross spectra.

Numerical generation has been employed in numerous scientific and engineering problems over decades, such as the simulation of
	earthquake ground motions \cite{Chang1981,Gersch1985,Deodatis1988},
	ocean waves \cite{Spanos1981,Spanos1983,Samii1984},
	atmospheric variables (mainly wind velocity fluctuations \cite{Reed1983,Li1990,Li1993,Deodatis1996,DiPaola2001,DiPaola2008,Kareem2008,Krenk2011,Krenk2019}, but also pressure \cite{Reed1983}, temperature and precipitation \cite{Sparks2018}, and variables for hydrological modelling like discharge and flood frequency \cite{Beven2021}),
	random vibration systems in the context of structural system identification   \cite{Gersch1972,Gersch1974,Gersch1976,Gersch1977}
	and spatial structures of several geological phenomena \cite{Sharifzadehlari2018,SoltanMohammadi2020}.

While many of the aforementioned real-life processes can only be rigourously represented through a non-stationary and/or non-homogeneous random process, stationarity and homogeneity are usually assumed in the modelling for convenience. In some cases, the obtained algorithms have served as a basis for the development of strategies oriented to non-stationary random processes, as in the case of evolutionary spectra, see the seminal paper \cite{Deodatis1988}.

There are different types of numerical generation approaches. Although an agreed classification based on common names for the different models lacks, see \cite{Kleinhans2009,Liu2019}, two families, referred to as spectral and sequential methods, are usually identified.
Spectral methods are based on strategies like harmonic superposition and inverse Fast Fourier Transform (FFT) \cite{Shinozuka1991,Shinozuka1996}. These methods require information regarding the spectral characterisation of the random process as an input, for example, in the form of predefined target cross power spectral density (CPSD) functions, coherence functions, etc. Some limitations of spectral methods are related to the fact that the process realisation needs to be synthesised in the whole time/space domain at once, which translates into high computational requirements for long-duration/long-distance multivariate and/or multi-dimensional processes \cite{Kareem2008}.

Sequential methods, also referred to as digital filters, are usually based on time series linear models, such as autoregressive (AR) and Moving Average (MA) models, or a combination of both (ARMA), and their multivariate versions (VAR, VMA and VARMA, respectively). 
Compared to spectral methods, sequential methods are less intensive in computational requirements during the synthesis, as only the model coefficients need to be stored. 
In addition, synthesis is a sequential process that can be stopped at a desired length of the time series and restarted later to lengthen the simulation. 
However, the determination of the model coefficients may demand high computational memory for multi-dimensional and/or multivariate problems. Model coefficients can be derived from a predefined target autocovariance function defined in time and/or space\footnote{For simplicity, we will refer only to univariate time series from here on, unless stated otherwise}, though some approaches introduce spectral information as input as well, see \cite{Spanos1981,Spanos1983}. 

Another advantage of sequential methods is that AR, MA, and ARMA models have theoretical expressions of their autocovariance and PSD functions that can be computed directly from the model coefficients. This fact allows comparing directly the target autocovariance function with the model theoretical autocovariance function to assess the accuracy of the model in reproducing the desired statistical information, while this comparison for the spectral methods is based on sample functions estimated from a finite number of finite-length realisations, subjected to smearing and leakage effects \cite{Stoica2005}. It is remarked that statistical bias affecting the sample autocovariance and PSD functions estimated from time series has to be taken into account when generating synthetic time series, regardless the nature of the method employed (spectral/sequential), in order to properly set the parameters of the simulations, like the time series length. \cite{Dimitriadis2015} provides expressions for the bias of estimators for both the PSD and autocovariance function, and discusses the advantages of the climacogram as an alternative statistical object to characterise random processes.

It is noted that, in the context of stationary random processes, both spectral and sequential methods can be applied regardless the form of the input information, since the power spectrum and the autocovariance function are Fourier pairs, thus they are two forms of providing the same statistical information. However, going from a PSD to an autocovariance function (and vice versa), except for particular cases that admit a theoretical formulation, requires a numerical implementation of the corresponding Fourier transform. For this reason, spectral methods are usually applied to problems where the input information is provided in the frequency domain, while time domain descriptions of a random process represent natural inputs for sequential methods.

This tutorial is focused on the determination of AR models to optimally reproduce a predefined target autocovariance function. Indeed, \textit{optimal} is a notion that needs to be clearly defined, as it will be discussed in Section \ref{sec_40_GAs}. To frame the problem, consider a one-dimensional univariate discrete random process, $\{z_t(\alpha)\}$, where $t \in \mathbb{N}$ is a  time index and $\alpha \in \mathbb{N}$ is the index of the realisation. Thus, $z_{t_i}(\alpha)$ is a random variable associated to time index $t_i$, and $\{z_t(\alpha_j)\}$ is a time series corresponding to the $\alpha_j$-th realisation of the random process. To simplify notation, the realisation index $\alpha$ will be omitted, so that $\{z_t\}$ will be used to refer a time series for a generic realisation $\alpha$, and $z_t$ to refer the random variable associated to time $t$. The random process is assumed to be Gaussian, stationary, and zero-mean. In addition, the existence of the integral time scale (i.e. the integral of the autocorrelation function) is assumed. Consequently, long-term persistence processes (Hurst phenomenon), characterised by an infinite integral scale, are excluded. The reason for this hypothesis is that this research considers stationary AR models, whose integral time scale always exists because the autocorrelation function decays exponentially. Long memory processes can be handled through different model types, like Fractional Autoregressive-Moving Average (FARMA) models \cite{Hipel1994}.

The general formulation of an AR model of order $p$, AR($p$), is as follows:

\begin{equation}\label{eq_AR_zeromean}
    z_t =  \sum_{i=1}^p \varphi_i \, z_{t-i} + \sigma \, \varepsilon_t,
\end{equation}

\noindent where $\varphi_i$ for $i=1,...,p$ are the regression coefficients of the AR model, and $\sigma \, \varepsilon_t$ represents the random term; $\varepsilon_t$ is a sequence of independent and identically distributed (iid) random variables with  zero mean and unit variance, and $\sigma$, here referred to as the noise coefficient, scales the variance of the random term, given by $\Var [\sigma \, \varepsilon_t] = \sigma^2$. Thus, the AR($p$) model has $p+1$ parameters, comprising $p$ regression coefficients and one noise coefficient.

\begin{figure}[ht!]
    \centering
    \includegraphics[width=0.75\textwidth]{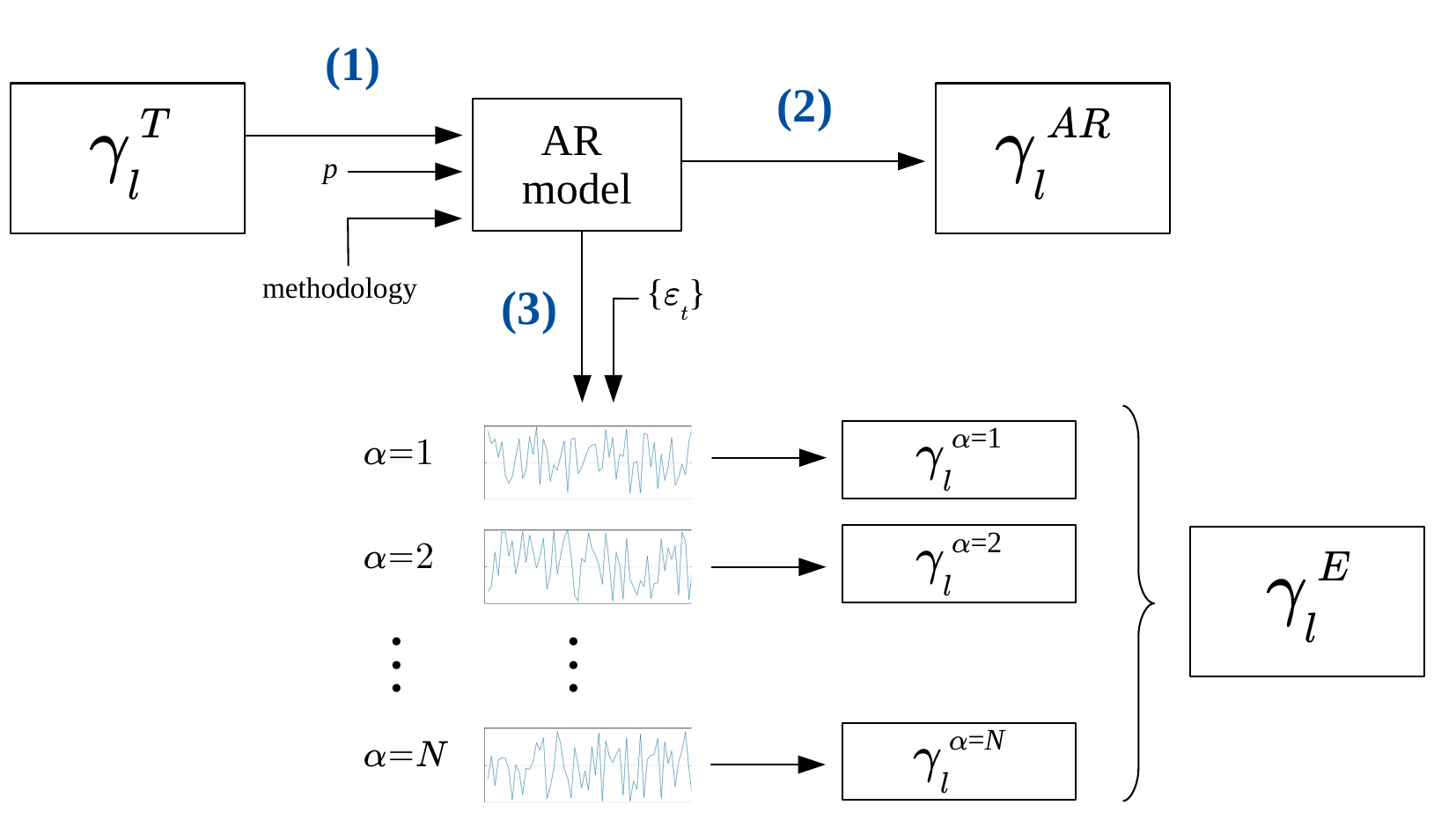}
    \caption{Scheme with the different elements involved in the synthetic generation of a random process with an AR model.}
    \label{fig_scheme}
\end{figure}

Figure \ref{fig_scheme} illustrates the different elements involved in the synthetic generation of a random process with an AR model (also valid for MA and ARMA models). 
$\gamma^T_{l}$ is the target autocovariance funcion, which depends only on the time lag $l$ under the assumption of stationarity (the formal definition of the autocovariance function is provided in Section \ref{sec_AR}). In some applications, $\gamma^T_{l}$ is derived from theoretical models and admits a mathematical expression, but in real-life problems it usually has to be estimated from observations. Step \textbf{(1)} represents the determination of the AR coefficients from $\gamma^T_{l}$; this step requires a methodology and a choice for the model order, $p$.
$\gamma^{AR}_{l}$ is the theoretical autocovariance function, and it can be computed directly from the AR model coefficients, step \textbf{(2)} in the figure.
Step \textbf{(3)} represents the synthesis of the random process through the AR model. By using a sequence of random values as inputs, $\{\varepsilon_t\}$, the AR model can be employed to generate realisations of the process in the form of time series. 
$\gamma^{\alpha}_{l}$ denotes the sample autocovariance function computed for realisation $\alpha$. 
Averaging $N$ sample autocovariance functions yields the ensemble autocovariance function, $\gamma^E_{l}$. $\gamma^E_{l}$ converges to $\gamma^{AR}_{l}$ when $N \longrightarrow \infty$. Thus, the objective in step \textbf{(1)} is to define a methodology that yields an AR model with a theoretical autocovariance function, $\gamma^{AR}_{l}$, that optimally reproduces the target, $\gamma^{T}_{l}$; this can be verified without the need for generating a large number of realisations $N$ to compute $\gamma^E_{l}$.

Regarding the methodologies to obtain the AR($p$) model parameters, the vast majority of works make use of the Yule-Walker (Y-W) equations \cite{Spanos1981,Spanos1983,Reed1983,Samaras1985}, which establish relationships between the $p+1$ AR parameters and the $p+1$ first autocovariance terms, from $l=0$ to $l=p$. 
These relationships arise simply from applying the definition of the autocovariance function for time lags from $0$ to $p$.
In the context of this work, these expressions are referred to as autocovariance equations for time lags from $0$ to $p$.
This approach leads to a perfect match of the first $p+1$ terms of the target autocovariance function. Consequently, under this approach, there exists no means to improve the matching between the target and the theoretical AR autocovariance functions for lags larger than the model order, $p$. 
This fact is problematic for processes with high inertia (i.e., large integral time scale) or situations in which small sampling times of the time series are employed, because large values of the model order $p$ are required to guarantee the matching of the target autocovariance function in a sufficiently wide range of time lags. But large $p$ values reduce model parsimony and increase the computational cost to determine the model coefficients \cite{Spanos1983,Dimitriadis2018}. Model parsimony refers to achieving a certain model performance with the lowest number of model parameters, and is considered a key feature in time series modelling \cite{Box2016}.
A proposal to preserve model parsimony is to use ARMA models \cite{Gersch1976,Samaras1985}. Several methodologies have been developed, typically based on multistage approaches that build the ARMA model by combining an AR($p$) model previously defined with a MA($q$) component. However, the potential of these approaches may be limited because MA processes show nonzero autocovariance only in the first $q+1$ terms, see for example \cite{Madsen2007}. Thus, the improvement of the matching between target and the theoretical autocovariance functions for large time lags could be conditioned to considering high $q$ values. In this article, a proposal for overcoming this limitation is introduced in Section \ref{subsec_fromVar2AR}. It consists in including autocovariance equations for lags larger than $p$ in the procedure.

In a recent paper \cite{Krenk2019}, an interesting proposal based on AR models only with regression coefficients at certain time lags was introduced for the synthetic generation of turbulent wind fields. The theoretical formulation is provided for a generic sequence of lags, and the simulations are performed for AR models with an exponential scheme (regression coefficients only for time lags $2^k$, $k=0,1,2,...$).
In econometrics, such models are usually referred to as \textit{restricted} AR models \cite{Giannini1987}, because they can be seen as a particular case of an AR($p$) model for which some of the regression coefficients are imposed to be zero.\footnote{In line with the literature, AR models refer to \textit{unrestricted} AR models, unless stated otherwise.} Thus, the actual number of regression coefficients is lower than the AR order, and the capacity of matching exactly the first $p+1$ terms of the target autocovariance function is lost. However, the trade-off between model parsimony and target autocovariance function reproducibility for a wide range of lags was improved. In our opinion, one limitation of that work is that the exponential scheme of the model  was assumed to be reasonably good for the considered application, and no further discussion is provided concerning the impact of different model schemes on the results. In what follows, the methodology presented in \cite{Krenk2019} will be referred to as the K-M approach.

Within this context, this article introduces a general formulation to determine the parameters of a restricted AR model from a predefined target autocovariance function. Under this general framework, it will be shown that both the Y-W approach and the K-M approach could be seen as particular cases of the presented formulation. 
The described approach requires a reduced number of input parameters related to the model scheme and the employed autocovariance equations. An optimisation procedure based on genetic algorithms is applied to obtain AR models that reproduce the target autocovariance function more accurately than the Y-W and K-M approaches for the same model parsimony. The main ideas contained in this tutorial and its research contributions are as follows:

\begin{itemize}
\item The Yule-Walker approach to obtain an AR($p$) model from a target autocovariance function is described, emphasising the classical result consisting in the perfect matching of the first $p+1$ terms of the target autocovariance function as a consequence of selecting a set of autocovariance equations for time lags from $0$ to $p$. 
\item We show that using autocovariance equations for time lags larger than $p$ may improve the matching of the target autocovariance function for lags beyond the model order.
\item The potential of restricted AR models for improving the matching of a target autocovariance function is revisited by considering the K-M approach introduced in \cite{Krenk2019}.
\item We introduce a general formulation for the AR parameters determination from a target autocovariance function. The formulation is general in the sense that it provides flexibility in the choice for the autocovariance equations and in the
definition of the AR model scheme.
\item The introduced formulation is exploited by a genetic algorithm to obtain optimal AR models without a pre-defined model scheme. The considered application is based on a stationary, homogeneous, and isotropic (SHI) turbulence model. Results are compared to those obtained with the Y-W approach and the K-M approach.
\item  The introduced general formulation is extended to the multivariate case. This leads to some computational bottlenecks that are highlighted.
\end{itemize}

The article is organised as follows. Section \ref{sec_AR} describes the relationships between the parameters of an AR model and its theoretical autocovariance function, emphasising the impact of the selected autocovariance equations on the matching of the target autocovariance function. The case of restricted AR models is addressed in Section \ref{sec_AR_restricted}, where the focus is placed on the role of the model scheme. The general formulation for the determination of a restricted AR model from a predefined target autocovariance function is introduced in Section \ref{sec_general_formulation}. Section \ref{sec_40_GAs} contains the optimisation exercise based on genetic algorithms. The generalisation of the problem to the mulivariate case is briefly described in Section \ref{sec_40_VARs}. The paper ends with the main conclusions gathered in Section \ref{sec_50_conclusions}. The article includes a number of examples and reflections to facilitate comprehension.



\section{The autocovariance equations of an AR model} \label{sec_AR} 

In time series analysis, the covariance between two random variables $z_{t_1}$ and $z_{t_2}$ is usually denoted by $\gamma_{t_1,t_2}$. Under the assumption of stationarity, the autocovariance depends only on the time lag, $l=t_1-t_2$, and is referred to as the autocovariance function:

\begin{equation} \label{eq_variance_definition}
     \gamma_l = \Cov [ \, z_t , z_{t-l} \, ] = \E[ \, z_t \, z_{t-l} \, ].
\end{equation}

Note that the autocovariance function is symmetric, $\gamma_{-l} = \gamma_l$. 
Note also that, since the random term in \eqref{eq_AR_zeromean} is independent, there is no dependency between the random term at time $t$, $\sigma \, \varepsilon_t$, and previous values of the process, $z_{t-l}$ for $l>0$. Indeed, the following expression can be demonstrated \cite{Madsen2007}:

\begin{equation} \label{eq_uncorr}
\E[ \, (\sigma \, \varepsilon_t) \,  z_{t-l} \, ] = \begin{cases}  \sigma^2 \text{~ for~} l=0 \\  0 \text{~ for~} l>0 \end{cases}.     
\end{equation}

Given these considerations, Equation (\ref{eq_variance_definition}) together with \eqref{eq_AR_zeromean} and \eqref{eq_uncorr} provide a means to generate analytical expressions that relate the autocovariance function for different time lags and the AR model parameters. As mentioned above, these expressions are here referred to as autocovariance equations.
The autocovariance equation for lag $l=0$ is:


\begin{equation} \label{eq_AR_variance_0}
\gamma_{0} 
= \Cov [ \, z_t , z_t \, ] =  \E[ \, z_t \, z_t \, ]
= \E \left[ \, \left( \sum_{i=1}^p \varphi_i \, z_{t-i} + \sigma \, \varepsilon_t \right) \, z_t \,  \right]
= \sum_{i=1}^p \varphi_i \, \gamma_{-i} + \sigma^2.
\end{equation}

The autocovariance equation for a generic positive time lag $l$ is:


\begin{equation} \label{eq_AR_variances}
\gamma_l 
= \Cov [ \, z_t , z_{t-l} \, ] 
= \E[ \, z_t \, z_{t-l} \, ]  
= \E \left[ \, \left( \sum_{i=1}^p \varphi_i \, z_{t-i} + \sigma \, \varepsilon_t \right) \, z_{t-l} \,  \right]
= \sum_{i=1}^p \varphi_i \, \gamma_{l-i}.  
\end{equation}

Note that Equation \eqref{eq_AR_variance_0} is the only one among all autocovariance equations that includes the noise coefficient, $\sigma$. Actually, this equation defines the relationship between the variance of the AR process, $\gamma_{0}$, and the variance of the random term, $\sigma^2$. 

Equations \eqref{eq_AR_variance_0} and \eqref{eq_AR_variances} are the basis for computing the theoretical autocovariance function of a given AR model, addressed in Section \ref{subsec_fromAR2Var}, and for obtaining the AR model parameters from a predefined target autocovariance function, see Section \ref{subsec_fromVar2AR}.

\subsection{Computing the theoretical autocovariance function of an AR model} \label{subsec_fromAR2Var}  

The objective of this section is to compute the first $n+1$ terms of the theoretical autocovariance function of an AR($p$) model, $\gamma_0^{AR}, \gamma_1^{AR},..., \gamma_{n}^{AR}$, assuming that the AR parameters, $\varphi_i$ for $i=1,...,p$ and $\sigma$, are known.

Without loss of generality, the case of an AR($2$) model is considered.
The following expression represents the autocovariance equations for lags from $l=0$ to $l=n$, see equations (\ref{eq_AR_variance_0}) and (\ref{eq_AR_variances}), in the form of a matrix equation, where the autocovariance terms have been gathered into the independent vector. 

\begin{equation}\label{eq_fromAR2Vars}
\begin{matrix}
l=0: \text{~~~} \\
l=1: \text{~~~} \\
l=2: \text{~~~} \\
l=3: \text{~~~} \\
\vdots   \text{~~~} \\
l=n: \text{~~~} \\
\end{matrix}
    \begin{pmatrix}
         -\varphi_2     & -\varphi_1     &  1            &  0               & 0            & 0          & \dots & 0  \\
        0               & -\varphi_2     & -\varphi_1    &  1               & 0            & 0          & \dots & 0	\\
        0               &  0             & -\varphi_2    & -\varphi_1       & 1            & 0          & \dots & 0  \\
        0               &  0             & 0             & -\varphi_2       & -\varphi_1   & 1          & \dots & 0  \\
        \vdots          &  \vdots        & \vdots        & \vdots           & \vdots       & \vdots     & \ddots & \vdots \\
        0               &  0             & 0             & 0                & 0	     & 0 	   & \dots & 1  \\
    \end{pmatrix} 
    \begin{pmatrix}
    \gamma_{-2} \\ \gamma_{-1} \\ \gamma_{0} \\ \gamma_{1} \\ \gamma_{2} \\ \gamma_{3} \\ \vdots \\ \gamma_{n}
    \end{pmatrix} 
    = 
    \begin{pmatrix}
    \sigma^2 \\ 0 \\ 0 \\ 0 \\ \vdots \\ 0 
    \end{pmatrix} .    
\end{equation}

Equation \eqref{eq_fromAR2Vars} represents a linear system of $n+1$ equations with $n+3$ unknowns.\footnote{~In the general case of an AR($p$) model, $n+1$ equations with $n+p+1$ unknowns.}
However, by applying symmetry in the autocovariance function, $\gamma_{-l}=\gamma_l$, it is possible to remove terms $\gamma_{-1}$ and $\gamma_{-2}$ from the independent vector. This allows one expressing the system of equations \eqref{eq_fromAR2Vars} with as many equations as unknowns:

\begin{equation}\label{eq_fromAR2Vars_v0}
\begin{matrix}
l=0: \text{~~~} \\
l=1: \text{~~~} \\
l=2: \text{~~~} \\
l=3: \text{~~~} \\
\vdots   \text{~~~} \\
l=n: \text{~~~} \\
\end{matrix}
    \begin{pmatrix}
        1             &  -\varphi_1      & -\varphi_2   & 0            & \dots  & 0 \\
        -\varphi_1    &  1-\varphi_2     & 0            & 0            & \dots  & 0 \\
        -\varphi_2    & -\varphi_1       & 1            & 0            & \dots  & 0 \\
        0             & -\varphi_2       & -\varphi_1   & 1            & \dots  & 0 \\
        \vdots        &  \vdots          & \vdots       & \vdots       & \ddots & \vdots \\
        0             & 0                & 0	         & 0 		 & \dots  & 1  \\
    \end{pmatrix}
     \begin{pmatrix}
    \gamma_{0} \\ \gamma_{1} \\ \gamma_{2} \\ \gamma_{3} \\ \vdots \\ \gamma_{n}
    \end{pmatrix} 
    =
    \begin{pmatrix}
    \sigma^2 \\ 0 \\ 0 \\ 0 \\ \vdots \\ 0
    \end{pmatrix} .    
\end{equation}

From that, the following expression for the theoretical autocovariance function of the AR(2) model is readily obtained:

\begin{equation}\label{eq_fromAR2Vars_v2}
    \begin{pmatrix}
    \gamma_{0}^{AR} \\ \gamma_{1}^{AR} \\ \gamma_{2}^{AR} \\ \gamma_{3}^{AR} \\ \vdots \\ \gamma_{n}^{AR}
    \end{pmatrix} 
    = 
    \begin{pmatrix}
        1             &  -\varphi_1      & -\varphi_2   & 0            & \dots  & 0 \\
        -\varphi_1    &  1-\varphi_2     & 0            & 0            & \dots  & 0 \\
        -\varphi_2    & -\varphi_1       & 1            & 0            & \dots  & 0 \\
        0             & -\varphi_2       & -\varphi_1   & 1            & \dots  & 0 \\
        \vdots        &  \vdots          & \vdots       & \vdots       & \ddots & \vdots \\
        0             & 0                & 0	         & 0 		 & \dots  & 1  \\
    \end{pmatrix} ^{-1}
    \begin{pmatrix}
    \sigma^2 \\ 0 \\ 0 \\ 0  \\ \vdots \\ 0
    \end{pmatrix} .    
\end{equation}

Equation (\ref{eq_fromAR2Vars_v2}) can be employed to obtain an arbitrary number $n$ of terms of $\gamma_{l}^{AR}$. Increasing $n$ comes at the expense of increasing the dimension of the matrix to be inverted, thus, the computational memory requirements. 
An alternative approach for computational alleviation consists in solving the subsystem given by the $p+1$ first autocovariance equations in order to obtain $\gamma_{0}^{AR}$, ..., $\gamma_{p}^{AR}$,

\begin{equation}\label{eq_fromAR2Vars_v3}
    \begin{pmatrix}
    \gamma_{0}^{AR} \\ \gamma_{1}^{AR} \\ \gamma_{2}^{AR} 
    \end{pmatrix} 
    = 
        \begin{pmatrix}
        1            &  -\varphi_1      & -\varphi_2   \\
        -\varphi_1    &  1-\varphi_2     & 0           \\
        -\varphi_2    & -\varphi_1       & 1           \\
    \end{pmatrix} ^{-1}
    \begin{pmatrix}
    \sigma^2 \\ 0 \\ 0 
    \end{pmatrix} ,
\end{equation}

\noindent and then to compute recursively $\gamma_{l}^{AR}$ for $l>p$ through Equation (\ref{eq_AR_variances}). The counterpart of this approach is that the accumulation of rounding errors may lead to inaccurate estimations of $\gamma_{l}^{AR}$ for large $l$ values.

\begin{figure}[ht]
    \centering
    \includegraphics[width=0.75\textwidth]{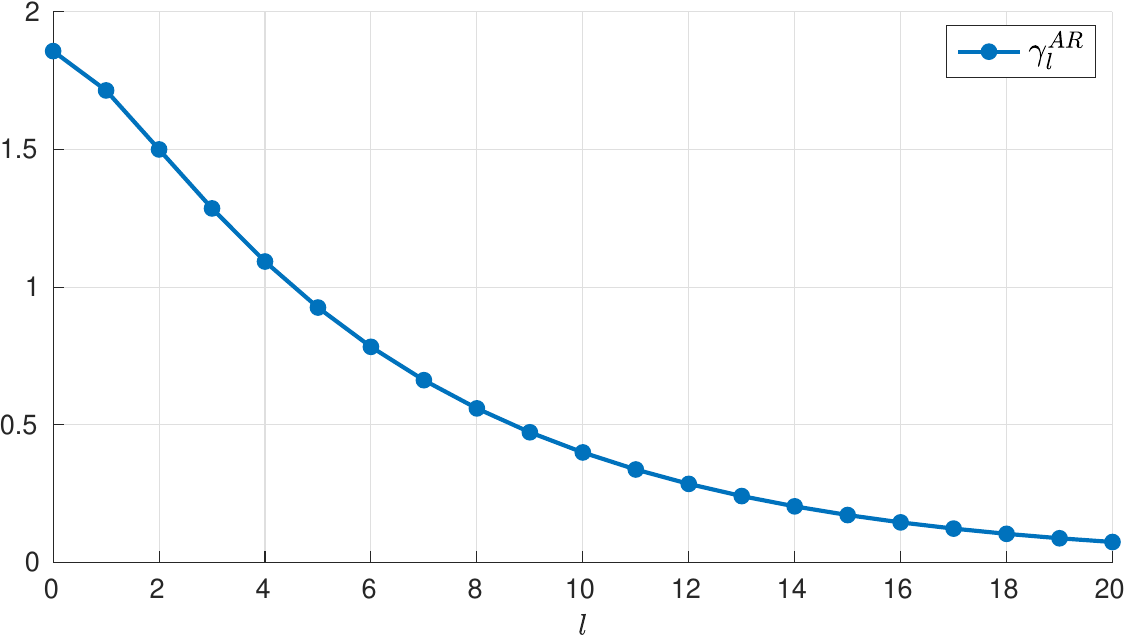}
    \caption{Theoretical autocovariance function of a predefined AR(2) model, $\varphi_1 = 1.2$, $\varphi_2 = -0.3$ and $\sigma = 0.5$.}
    \label{fig_fromAR2Var}
\end{figure}

As an example, Figure \ref{fig_fromAR2Var} shows $\gamma_{l}^{AR}$ for an AR(2) model given by $\varphi_1 = 1.2$, $\varphi_2 = -0.3$ and $\sigma = 0.5$, computed for lags up to $l=20$.

\subsection{Computing the AR model parameters from a predefined target autocovariance function} \label{subsec_fromVar2AR} 

The objective of this section is to compute the parameters of an AR model from a predefined target autocovariance function, $\gamma^T_l$, assuming that the target is available for any time lag $l$. It will be shown that only a limited number of values of $\gamma^T_l$ are required, depending on the employed autocovariance equations.

The $p+1$ model parameters, $\varphi_1$, $\varphi_2$, ..., $\varphi_p$ and  $\sigma$, are computed from $p+1$ autocovariance equations. 
The traditional approach to this problem considers the autocovariance equations for lags $l=0,1,...,p$, which leads to the Yule-Walker (Y-W) equations. For this reason, this approach is here referred to as the Y-W approach. For illustrative purposes, consider the case of an AR($3$) model, note that symmetry in the autocovariance function has already been applied:

\begin{align} \label{eq_AR3_variances_system}
    l=0 : ~~ \gamma_{0} & = \varphi_1 \, \gamma_{1} + \varphi_2 \, \gamma_{2} + \varphi_3 \, \gamma_{3} + \sigma^2 ,  \nonumber \\
    l=1 : ~~ \gamma_{1} & = \varphi_1 \, \gamma_{0} + \varphi_2 \, \gamma_{1} + \varphi_3 \, \gamma_{2} ,               \\
    l=2 : ~~ \gamma_{2} & = \varphi_1 \, \gamma_{1} + \varphi_2 \, \gamma_{0} + \varphi_3 \, \gamma_{1} ,              \nonumber \\
    l=3 : ~~ \gamma_{3} & = \varphi_1 \, \gamma_{2} + \varphi_2 \, \gamma_{1} + \varphi_3 \, \gamma_{0} .              \nonumber 
\end{align}

Equation \eqref{eq_AR3_variances_system} can be written in matrix form as:

\begin{equation} \label{eq_AR3_solution_0123}
    \begin{pmatrix}
    \gamma_{0} \\ \gamma_{1} \\ \gamma_{2} \\ \gamma_{3}
    \end{pmatrix} 
    =
    \begin{pmatrix}
        1	&	\gamma_{1}	&   \gamma_{2}	&   \gamma_{3} \\
        0	&	\gamma_{0}	&   \gamma_{1}	&   \gamma_{2}	\\
        0 	&	\gamma_{1}	&   \gamma_{0}	&   \gamma_{1}	\\
        0 	&	\gamma_{2}	&   \gamma_{1}	&   \gamma_{0}	\\
        \end{pmatrix}
	\,
    \begin{pmatrix}
      \sigma^2  \\  \varphi_1  \\  \varphi_2 \\ \varphi_3 
    \end{pmatrix}.
\end{equation}

By replacing in \eqref{eq_AR3_solution_0123} the autocovariance terms $\gamma_i$ by the corresponding target values, $\gamma^T_i$, the four AR model parameters are obtained from $\gamma^T_{0}, ... ,\gamma^T_{3}$:

\begin{equation} \label{eq_AR3_solution_0123_rearranged}
    \begin{pmatrix}
      \sigma^2  \\  \varphi_1  \\  \varphi_2 \\ \varphi_3 
    \end{pmatrix}
	=
    \begin{pmatrix}
        1	&	\gamma^T_{1}	&   \gamma^T_{2}	&   \gamma^T_{3} \\
        0	&	\gamma^T_{0}	&   \gamma^T_{1}	&   \gamma^T_{2}	\\
        0 	&	\gamma^T_{1}	&   \gamma^T_{0}	&   \gamma^T_{1}	\\
        0 	&	\gamma^T_{2}	&   \gamma^T_{1}	&   \gamma^T_{0}	\\
        \end{pmatrix}^{-1}
	\,
\begin{pmatrix}
    \gamma^T_{0} \\ \gamma^T_{1} \\ \gamma^T_{2} \\ \gamma^T_{3}
    \end{pmatrix}.
\end{equation}

A key consequence of employing the autocovariance equations for lags $l=0,...,3$ is that number of model parameters and the number of required $\gamma^T_l$ values is the same, which leads to an AR(3) model with a theoretical autocovariance function that matches exactly the employed target values. This conclusion can be extended for an AR($p$) and the first $p+1$ terms of the target autocovariance function. 

As an example, the following AR(3) model has been obtained for the target autocovariance function described in Appendix \ref{A_isotropic}:

\begin{equation} \label{ex_model_YW}
    z_t = 0.663 \, z_{t-1} + 0.099 \, z_{t-2} + 0.044 \, z_{t-3} + 0.636 \, \varepsilon_t.
\end{equation}

\begin{figure}[ht]
    \centering
    \includegraphics[width=0.75\textwidth]{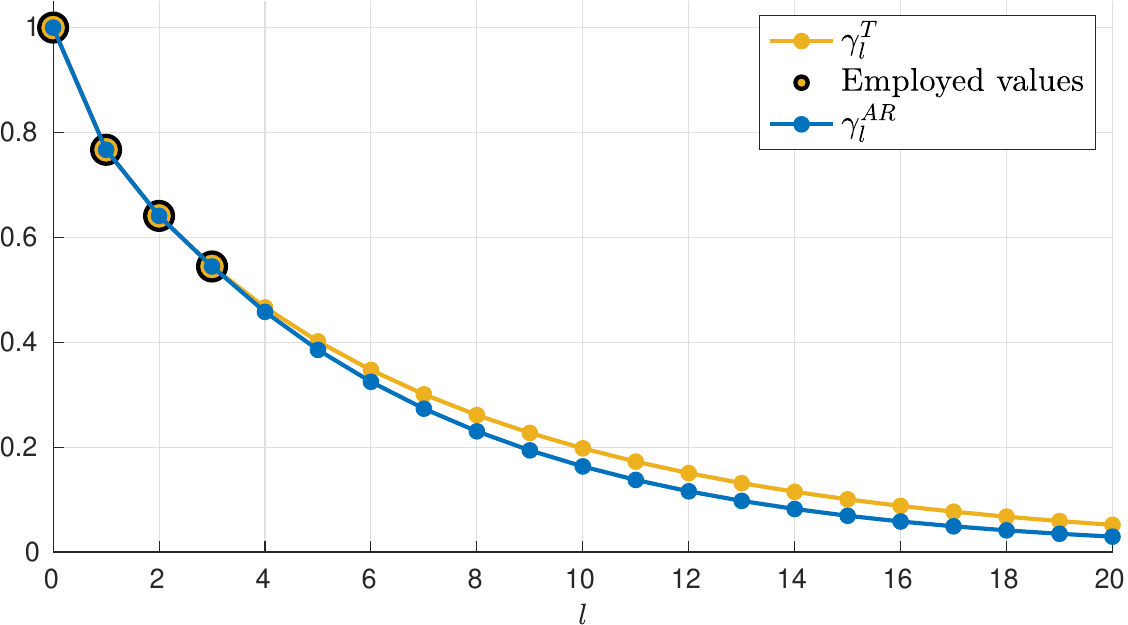}
    \caption{Autocovariance function of an AR(3) model and target autocovariance function. The AR model was obtained by considering the autocovariance equations for lags $l=0,1,2,3$. The employed values of the target autocovariance function are highlighted.}
    \label{fig_AR3_0123}
\end{figure}

The theoretical autocovariance function of the obtained AR model, $\gamma^{AR}_\tau$, and the target autocovariance function, $\gamma^{T}_\tau$, are shown in Figure \ref{fig_AR3_0123}. The four target values employed during the model determination have been highlighted. Note that the theoretical autocovariance function of the AR(3) model matches exactly the employed target autocovariance values, but increasing differences with the target are observed for larger lags.

Note also the following comments:

\begin{itemize} 
    \item [(i)] The determination of the AR model involves a matrix inversion. Care should be taken with issues related to ill-conditioned matrices that may arise from some target autocovariance functions defined arbitrarily. For example, target values $\gamma^T_{0}=1$, $\gamma^T_{1}=0.5$ and $\gamma^T_{2}=-0.5$ lead to a non-invertible matrix in \eqref{eq_AR3_solution_0123_rearranged}, regardless the value of $\gamma^T_3$.
    \item [(ii)] While the autocovariance function of the obtained AR model reproduces exactly $\gamma^T_{l}$ for $l=0,...,p$, no constraints have been imposed on the autocovariance function for larger time lags $l > p$. This implies that there is no means to improve the matching between $\gamma^{AR}_{l}$ and $\gamma^T_{l}$ for time lags larger than $p$. 
\end{itemize}

The last comment leads to an important question: is it possible to introduce information of $\gamma^T_{l}$ for time lags larger than the AR model order in the determination of the model parameters? If so, that would provide a means to obtain AR($p$) models for which there exists some control on $\gamma^{AR}_{l}$ for $l>p$, potentially improving the trade-off between model parsimony and matching between target and theoretical autocovariance function, compared to the Y-W approach. 
To address this question, autocovariance equations for time lags larger than $p$ could be considered. Let us define vector $\textbf{l}=[l_1,l_2,...,l_N]$ with the $N$ positive lags corresponding to the autocovariance equations employed in the model determination. By default, the autocovariance equation for $l=0$ is always required to determine the noise parameter, $\sigma$. For this reason, only positive lags are specified in vector $\textbf{l}$. Note also that $N$ must be equal to the number of model regression coefficients, $p$. As an example, consider the set of autocovariance equations obtained with $\textbf{l} = [1, 2, 5]$ for an AR(3) model:

\begin{align} \label{eq_AR3_variances_system_0125}
    l=0:  ~~ \gamma_{0} & = \varphi_1 \, \gamma_{1} + \varphi_2 \, \gamma_{2} + \varphi_3 \, \gamma_{3} + \sigma^2   \nonumber \\
    l=1:  ~~ \gamma_{1} & = \varphi_1 \, \gamma_{0} + \varphi_2 \, \gamma_{1} + \varphi_3 \, \gamma_{2}               \\
    l=2:  ~~ \gamma_{2} & = \varphi_1 \, \gamma_{1} + \varphi_2 \, \gamma_{0} + \varphi_3 \, \gamma_{1}           \nonumber \\
    l=5:  ~~ \gamma_{5} & = \varphi_1 \, \gamma_{4} + \varphi_2 \, \gamma_{3} + \varphi_3 \, \gamma_{2}      .        \nonumber 
\end{align}

Now, the AR model parameters are given by:

\begin{equation} \label{eq_AR3_solution_0125}
    \begin{pmatrix}
      \sigma^2  \\  \varphi_1  \\  \varphi_2 \\ \varphi_3 
    \end{pmatrix}
	=
    \begin{pmatrix}
        1	&	\gamma^T_{1}	&   \gamma^T_{2}	&   \gamma^T_{3} \\
        0	&	\gamma^T_{0}	&   \gamma^T_{1}	&   \gamma^T_{2}	\\
        0 	&	\gamma^T_{1}	&   \gamma^T_{0}	&   \gamma^T_{1}	\\
        0 	&	\gamma^T_{4}	&   \gamma^T_{3}	&   \gamma^T_{2}	\\
        \end{pmatrix}^{-1}
	\,
\begin{pmatrix}
    \gamma^T_{0} \\ \gamma^T_{1} \\ \gamma^T_{2} \\ \gamma^T_{5}
    \end{pmatrix}.
\end{equation}

\begin{figure}[ht!]
    \centering
    \includegraphics[width=0.75\textwidth]{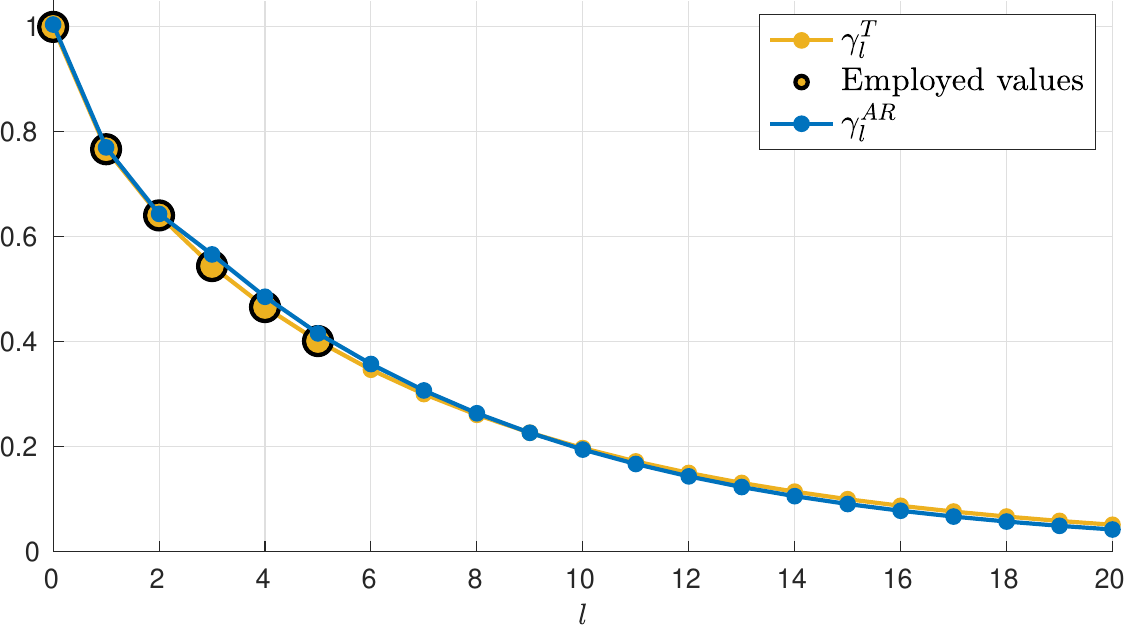}
    \caption{Autocovariance function of an AR(3) model and target autocovariance function. The AR model was obtained by considering the autocovariance equations for lags $l=0$ and $\textbf{l}=[1,2,5]$. The employed values of the target autocovariance function are highlighted.}
    \label{fig_AR3_0125}
\end{figure}

Equation  \eqref{eq_AR3_solution_0125} reveals that, in this case, the four AR model parameters are computed from six terms of the target autocovariance function, from $\gamma_0^T$ to $\gamma_5^T$.
For the particular case of the target autocovariance function described in Appendix \ref{A_isotropic}, the obtained AR(3) model is:

\begin{equation} \label{ex_model_l_vector}
    z_t = 0.657 \, z_{t-1} + 0.066 \, z_{t-2} + 0.092 \, z_{t-3} + 0.635 \, \varepsilon_t,
\end{equation}

\noindent which differs notably from the model obtained with the Y-W approach, see \eqref{ex_model_YW}.
Figure \ref{fig_AR3_0125} shows the target and the theoretical autocovariance functions, $\gamma_l^T$ and $\gamma_l^{AR}$, respectively. The six target values employed during the model determination have been highlighted. It can be seen that $\gamma_l^{AR}$ does not match exactly $\gamma_l^T$ for any time lag. However, a visual comparison with Figure \ref{fig_AR3_0123} reveals that the global matching is improved.

From this analysis, it can be concluded that, in the determination of the parameters of an AR$(p)$ model, using autocovariance equations for lags larger than $p$ leads to a number of required target terms higher than the number of model parameters. Since the autocovariance equations represent constraints between the AR model parameters and certain terms of the autocovariance function, this inequality makes that $\gamma_l^{AR}$ does not exactly match the target values employed in the model determination, Equation \eqref{eq_AR3_solution_0125}, as it was the case for the Y-W approach; $\gamma_l^{AR}$ only fulfils the constraints determined by the selected autocovariance equations. However, given a particular target $\gamma_l^{T}$ and a considered AR model order $p$, optimal decisions on the selected autocovariance equations (defined in vector $\textbf{l}$) may lead to AR models that reproduce $\gamma_l^{T}$ globally better, as compared with the models obtained with the Y-W approach, as it was shown in the previous example. Thus, considering vector $\textbf{l}$ as an input parameter in the determination of an AR($p$) model introduces flexibility as compared with the assumption $\textbf{l}=[1,...,p]$ that underlies the Y-W approach, and represents a path for improvement in reproducing a predefined target autocovariance function. To the authors knowledge, this strategy has not been addressed previously in the literature.


\section{The autocovariance equations of a restricted AR model} \label{sec_AR_restricted}

In a restricted AR model, not all regression coefficients from $\varphi_1$ to $\varphi_p$ are considered. Let us define vector $\textbf{j} = [ j_1 , j_2 , ... , j_N] $, with $N<p$, containing the lags of the regression terms included in the model. The AR order is given by the regression term with the highest lag, $p \equiv j_N$. Note that $N$ represents the number of regression coefficients of the model. To distinguish between restricted and unrestricted AR models, the following notation will be employed for restricted AR models:

\begin{equation}\label{eq_AR_restricted}
    z_t =  \sum_{i=1}^N a_{j_i} \, z_{t-j_i} + b \, \varepsilon_t,
\end{equation}

%

A restricted AR($p,\textbf{j}$) model can be seen as a particular case of an AR($p$) model for which:

\begin{equation} \label{u2r1}
    \varphi_{h} = \begin{cases}
 a_{h} \, , \text{~for~} h \in \textbf{j} \\
 0   ~~ , \text{~for~} h \notin \textbf{j} \\
 \end{cases} ,~ h=1,...,p.
\end{equation}

and 

\begin{equation} \label{u2r2}
\sigma = b .
\end{equation}

Note also that an AR($p$) can be seen as a particular case of a restricted AR($p,\textbf{j}$) model for which:

\begin{equation} \label{r2u}
\textbf{j}= [ \, 1, 2, 3, ..., p \,] ,
\end{equation}

\noindent provided that the constraint $N < p$ is relaxed.

The autocovariance equations of a restricted AR($p,\textbf{j}$) process can be readily obtained by combining the autocovariance equations of the corresponding unrestricted AR($p$) model, see Section \ref{sec_AR}, with equations \eqref{u2r1} and \eqref{u2r2}:

\begin{align} \label{eq_AR_variance_0_restricted}
    \gamma_{0} = \sum_{i=1}^N a_{j_i} \, \gamma_{-j_i} + b^2,
\end{align}

\begin{align} \label{eq_AR_variances_restricted}
    \gamma_{l} = \sum_{i=1}^N a_{j_i} \, \gamma_{l-j_i}.
\end{align}

\subsection{Computing the theoretical autocovariance function of a restricted AR model} \label{subsec_fromAR2Var_restricted} 

\begin{figure}[ht]
    \centering
    \includegraphics[width=0.75\textwidth]{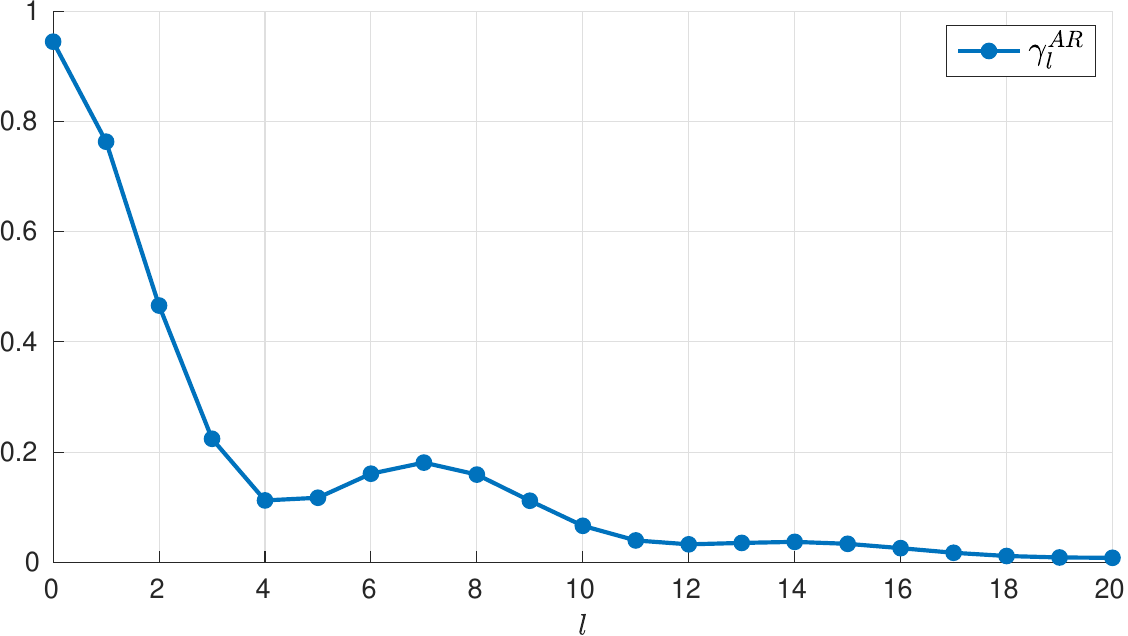}
    \caption{Autocovariance function of a predefined AR($5,\textbf{j}=[1,2,5]$) model with $a_1 = 1.2$, $a_2 = -0.5$, $a_5 = 0.1$  and $b = 0.5$.}
    \label{fig_fromAR2Var_restricted}
\end{figure}

The problem of computing $\gamma_l^{AR}$ for a restricted AR($p,\textbf{j}$) model can be readily addressed by applying the procedure described in Section \ref{subsec_fromAR2Var} together with eqs. \eqref{u2r1} and \eqref{u2r2}. 
As an example, Figure \ref{fig_fromAR2Var_restricted} shows $\gamma_l^{AR}$ for an AR($5,[1,2,5]$) model given by $a_1 = 1.2$, $a_2 = -0.5$, $a_5 = 0.1$  and $b = 0.5$,  computed for lags up to $l=20$.

A peculiarity of restricted AR models is that specific model schemes (i.e., specific values of vector $\textbf{j}$ components) lead to theoretical autocovariance functions with alternating zero and nonzero values. For example, let us consider the model AR(3,$[3]$):

\begin{equation}
    z_t = a_3  \, z_{t-3} + b \, \varepsilon_t
\end{equation}

The values of the autocovariance function $\gamma_l^{AR}$ up to lag $l=3$ are provided by the following autocovariance equations:

\begin{align} \label{eq_sis_ar3}
    l=0: & ~~ \gamma_{0} = a_3 \, \gamma_{-3} + b^2 ,\nonumber \\
    l=1: & ~~ \gamma_{1} = a_3 \, \gamma_{-2} ,\\
    l=2: & ~~ \gamma_{2} = a_3 \, \gamma_{-1} ,\nonumber \\
    l=3: & ~~ \gamma_{3} = a_3 \, \gamma_{0}  .\nonumber
\end{align}

By applying symmetry to the autocovariance function, the system of equations given in \eqref{eq_sis_ar3} can be expressed as follows:

\begin{equation} \label{eq_restricted_with_zeros}
    \begin{pmatrix}
        1       &   0       &   0       &   -a_3    \\
        0       &   1       &   -a_3    &   0       \\
        0       &   -a_3    &   1       &   0       \\
        -a_3    &   0       &   0       &   1   
    \end{pmatrix}
    \, 
    \begin{pmatrix}
        \gamma_{0}    \\
        \gamma_{1}    \\
        \gamma_{2}    \\
        \gamma_{3}    \\
    \end{pmatrix}
    =
        \begin{pmatrix}
    b^2 \\ 0 \\ 0 \\ 0 
    \end{pmatrix}.
\end{equation}

Equation \eqref{eq_restricted_with_zeros} can be divided into two independent subsystems. The first one with equations for lags $l=0$ and $l=3$, and the second one for lags for which there are no regression terms, $l=1$ and $l=2$. Solving the first subsystem yields:

\begin{equation} \label{ex_solution_AR_3_restricted}
    \gamma_{0}^T = \frac{b^2}{1-a_3^2}  ~~ \text{, and} ~~ \gamma_{3}^T = \frac{b^2}{1-a_3^2} \, a_3.    
\end{equation} 

Concerning the second subsystem, the only possible solution is:

$$ \gamma_{1}^T = \gamma_{2}^T = 0.$$

By recursively applying Equation \eqref{eq_AR_variances_restricted}, one gets nonzero values for $\gamma_l^{AR}$ only at lags $l=3,6,9,...$. In a general case, a restricted AR($p$) model with a single regression term at lag $\textbf{j}=[p]$ has a theoretical autocovariance function with non-zero values only at lags $l=p$ and its multiples.
This particular structure of $\gamma_l^{AR}$ is also observed for restricted AR models with $\textbf{j}$ vectors such that $j_i$ is multiple of $j_1$ for $i>1$. For example, the theoretical autocovariance function of an AR(6,$[2 , 4 , 6]$) model will show nonzero values only at even time lags. Such model schemes are likely to represent bad candidates for reproducing real-life autocovariance functions that usually fade out to zero in a continuous fashion.

\subsection{Computing the parameters of a restricted AR model from a predefined target autocovariance function} \label{subsec_fromVar2AR_restricted} 

In Section \ref{subsec_fromVar2AR} it was shown that the $p+1$ parameters of an (unrestricted) AR($p$) model can be computed to reproduce the first $p+1$ values of a predefined target autocovariance function through the Y-W approach. It was also discussed how the exact matching up to lag $p$ could be sacrificed in favour of improving the global matching by employing autocovariance equations for lags larger than $p$. In this section, restricted AR models are considered as an additional strategy to increase the control on the obtained model autocovariance function for time lags larger than the number of regression coefficients, $N$.

Without loss of generality, consider the problem of computing the parameters of a restricted AR model given by $\textbf{j}=[1,2,5]$. Note that the model parsimony is the same than that of the AR(3) model employed in Section \ref{subsec_fromVar2AR}, i.e. $N=3$, but in this case the model order is $p=5$. Let us start by considering the autocovariance equations of the corresponding unrestricted AR(5) model, for lags $l=0,...,5$:

\begin{align} \label{eq_AR5_variances_system}
    l=0: ~~ \gamma_{0} & = \varphi_1 \, \gamma_{1} + \varphi_2 \, \gamma_{2} + \varphi_3 \, \gamma_{3} + \varphi_4 \, \gamma_{4} + \varphi_5 \, \gamma_{5} + \sigma^2   , \nonumber \\
    l=1: ~~ \gamma_{1} & = \varphi_1 \, \gamma_{0} + \varphi_2 \, \gamma_{1} + \varphi_3 \, \gamma_{2} + \varphi_4 \, \gamma_{3} + \varphi_5 \, \gamma_{4} , \nonumber  \\
    l=2: ~~ \gamma_{2} & = \varphi_1 \, \gamma_{1} + \varphi_2 \, \gamma_{0} + \varphi_3 \, \gamma_{1} + \varphi_4 \, \gamma_{2} + \varphi_5 \, \gamma_{3}  , \\
    l=3: ~~ \gamma_{3} & = \varphi_1 \, \gamma_{2} + \varphi_2 \, \gamma_{1} + \varphi_3 \, \gamma_{0} + \varphi_4 \, \gamma_{1} + \varphi_5 \, \gamma_{2} , \nonumber \\
    l=4: ~~ \gamma_{4} & = \varphi_1 \, \gamma_{3} + \varphi_2 \, \gamma_{2} + \varphi_3 \, \gamma_{1} + \varphi_4 \, \gamma_{0} + \varphi_5 \, \gamma_{1} , \nonumber \\
    l=5: ~~ \gamma_{5} & = \varphi_1 \, \gamma_{4} + \varphi_2 \, \gamma_{3} + \varphi_3 \, \gamma_{2} + \varphi_4 \, \gamma_{1} + \varphi_5 \, \gamma_{0} . \nonumber 
\end{align}

The system of equations \eqref{eq_AR5_variances_system} contains six equations, six model parameters ($\sigma$ , $\varphi_1$, ..., $\varphi_5$) and six autocovariance terms ($\gamma_{0}, \gamma_{1}, ... \gamma_{5}$). Thus, according to the Y-W approach described in Section \ref{subsec_fromVar2AR}, by introducing the six target autocovariance terms, the system of equations is linear and provides the six parameters of an AR model whose theoretical autocovariance function matches exactly the imposed target autocovariance values.

Now, consider the restricted model AR(5,$[1,2,5]$), obtained from an AR(5) by imposing $\varphi_3=0$ and $\varphi_4=0$. Since these two model parameters are no longer unknowns in \eqref{eq_AR5_variances_system}, two possible strategies can be followed to obtain the model parameters ($\varphi_1$, $\varphi_2$, $\varphi_5$ and $\sigma$) from \eqref{eq_AR5_variances_system}: 

\begin{enumerate}
    \item \label{item_1} To select two new unknowns from the set of six autocovariance terms. These two terms will not be replaced by target autocovariance values. This yields a system of equations with six equations and six unknowns (four model parameters and the two selected autocovariance terms).
    \item \label{item_2} To discard two autocovariance equations, in order to have a system of equations with four equations and four unknowns (the four model parameters).
\end{enumerate}

If the strategy defined in \ref{item_1} is considered, only four (and not six) values of the target autocovariance function are required. Let us select, with no loss of generality, $\gamma_{0}$, $\gamma_{1}$, $\gamma_{3}$ and $\gamma_{5}$ to be replaced by the corresponding target autocovariance values, and consider $\gamma_{2}$ and $\gamma_{4}$ as additional unknowns. In this case, the system of equations \eqref{eq_AR5_variances_system} yields the six unknowns ($\sigma$, $\varphi_1$, $\varphi_2$, $\varphi_5$, $\gamma_{2}$ and $\gamma_{4}$). The drawback of this strategy is that the system of equations becomes non-linear, due to the products between unknowns such as $\varphi_2 \, \gamma_{2}$ and $\varphi_5 \, \gamma_{4}$ in the first and second equations, respectively. Note that the computational cost may be dramatically increased, specially for AR models with large order $p$, as the system of equations to be solved comprises $p+1$ equations, regardless the number of AR coefficients assigned to zero. The advantage of this strategy is that the obtained restricted AR model has a theoretical autocovariance function that matches exactly the four provided target autocovariance values.

To illustrate this, the restricted AR(5,$[1,2,5]$) model was obtained for the target autocovariance function described in Appendix \ref{A_isotropic}:

\begin{equation} \label{ex_AR125_0135_nonlinear}
    z_t = 0.649 \, z_{t-1} + 0.138 \, z_{t-2} + 0.026 \, z_{t-5} + 0.634 \, \varepsilon_t.
\end{equation}

\begin{figure}[ht]
    \centering
    \includegraphics[width=0.75\textwidth]{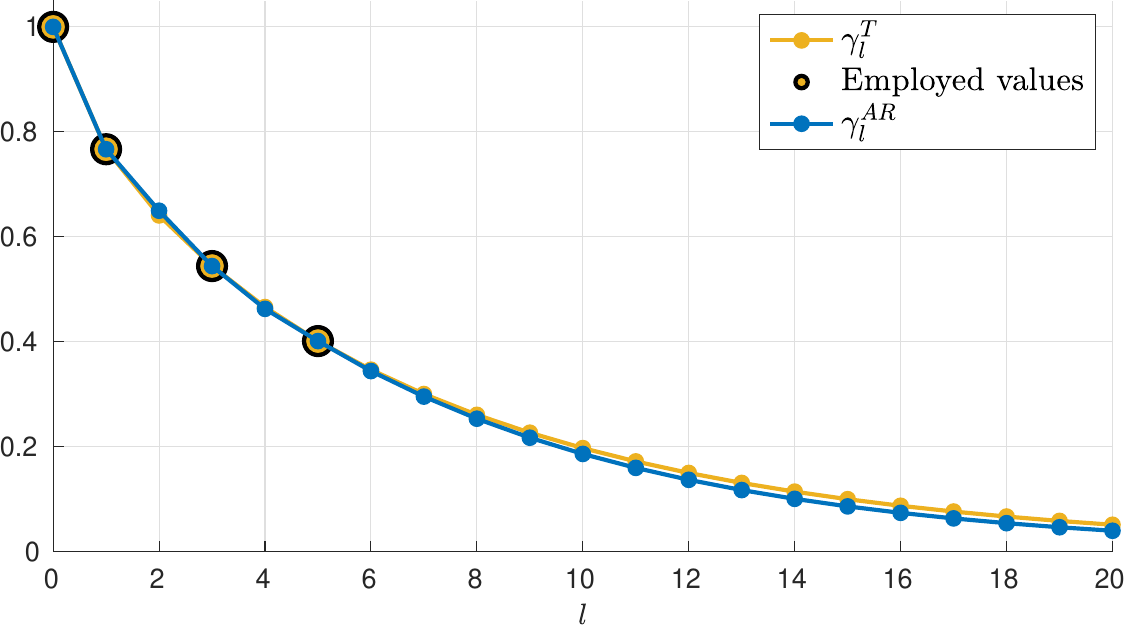}
    \caption{Autocovariance function of an AR(5,\textbf{j}=[1,2,5]) model and target autocovariance function. The restricted AR model was obtained with the non-linear approach, see text for details. The employed values of the target autocovariance function are highlighted.}
    \label{fig_AR125_0135_nonlinear}
\end{figure}

Figure \ref{fig_AR125_0135_nonlinear} shows the corresponding $\gamma_l^{AR}$ and $\gamma_l^{T}$ functions, where the four target values employed during the model determination have been highlighted. The figure shows an exact matching of $\gamma^T_{0}$, $\gamma^T_{1}$, $\gamma^T_{3}$ and $\gamma^T_{5}$, as well as an improved global matching of the target autocovariance function with respect to the Y-W approach for the same model parsimony can be observed, see Figure \ref{fig_AR3_0123}.

The second strategy defined in \ref{item_2} is actually equivalent to just selecting as many autocovariance equations as AR parameters, by defining a vector \textbf{l}. For the considered example, and without loss of generality, the autocovariance equations for lags $l=0$ and $\textbf{l}=[1,4,5]$ are selected. The resulting system of equations, using the notation for restricted AR models, is:

\begin{align} \label{eq_AR5_variances_system_reduced}
    l=0: ~~ \gamma_{0} & = a_1 \, \gamma_{1} + a_2 \, \gamma_{2} + a_5 \, \gamma_{5} + b^2 ,  \nonumber \\
    l=1: ~~ \gamma_{1} & = a_1 \, \gamma_{0} + a_2 \, \gamma_{1} + a_5 \, \gamma_{4}  , \\
    l=4: ~~ \gamma_{4} & = a_1 \, \gamma_{3} + a_2 \, \gamma_{2} + a_5 \, \gamma_{1} , \nonumber \\
    l=5: ~~ \gamma_{5} & = a_1 \, \gamma_{4} + a_2 \, \gamma_{3} + a_5 \, \gamma_{0} . \nonumber 
\end{align}

The system of equations \eqref{eq_AR5_variances_system_reduced} is linear, and comprises four equations, four unknowns (the model parameters $b^2$, $a_1$, $a_2$ and $a_5$), and requires six autocovariance terms. By introducing the target autocovariance terms, the model parameters are given by:

\begin{equation} \label{eq_AR5_variances_system_reduced_sol}
    \begin{pmatrix}
      b^2  \\  a_1  \\  a_2 \\ a_5 
    \end{pmatrix}
	=
    \begin{pmatrix}
        1	&	\gamma^T_{1}	&   \gamma^T_{2}	&   \gamma^T_{5} \\
        0	&	\gamma^T_{0}	&   \gamma^T_{1}	&   \gamma^T_{4}	\\
        0 	&	\gamma^T_{3}	&   \gamma^T_{2}	&   \gamma^T_{1}	\\
        0 	&	\gamma^T_{4}	&   \gamma^T_{3}	&   \gamma^T_{0}	\\
        \end{pmatrix}^{-1}
	\,
\begin{pmatrix}
    \gamma^T_{0} \\ \gamma^T_{1} \\ \gamma^T_{4} \\ \gamma^T_{5}
    \end{pmatrix}.
\end{equation}

\begin{figure}[ht!]
    \centering
    \includegraphics[width=0.75\textwidth]{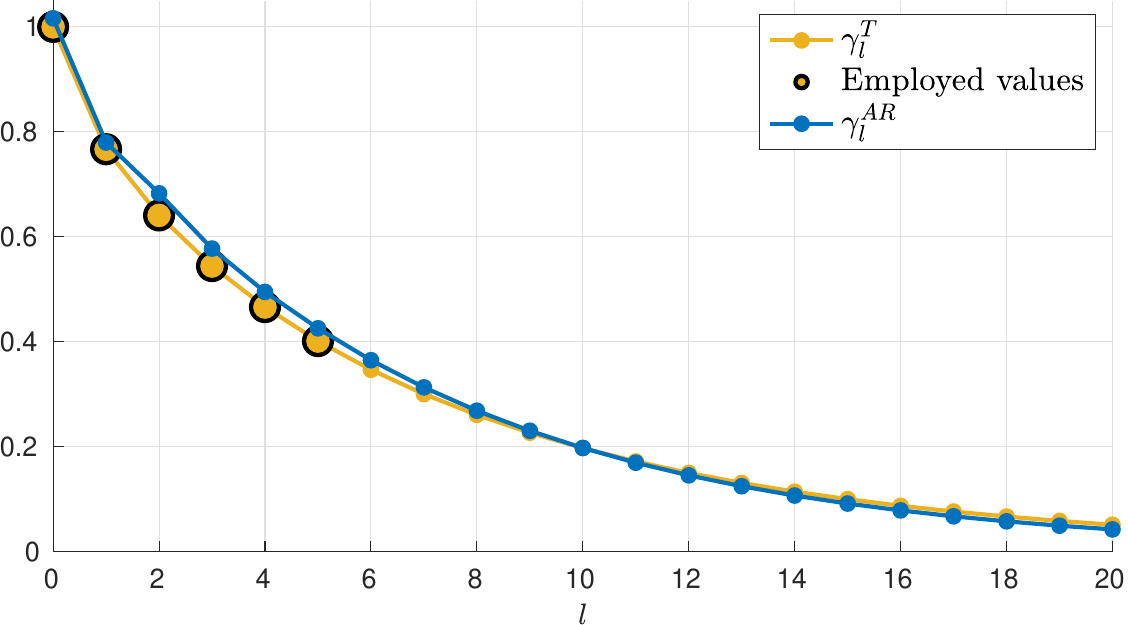}
    \caption{Autocovariance function of an AR(5,\textbf{j}=[1,2,5]) model and target autocovariance function. The restricted AR model was obtained by considering the autocovariance equations for lags $l=0$ and $\textbf{l}=[1,4,5]$. The employed values of the target autocovariance function are highlighted.}
    \label{fig_AR125_0145}
\end{figure}

This situation is similar to that explained in Section \ref{subsec_fromAR2Var}, in the sense that, since the number of required target autocovariance terms is higher than the number of equations (i.e., the model parameters), the autocovariance function of the obtained AR model will not exactly match any of the imposed target autocovariance values, but it may show a reasonably good matching for a wide range of time lags.
To illustrate this idea, Figure \ref{fig_AR125_0145} shows $\gamma_l^{AR}$ and $\gamma_l^{T}$ functions of the AR(5,[1,2,5]) model computed with $\textbf{l}=[1,4,5]$ for the target autocovariance function described in Appendix \ref{A_isotropic}. The six target values employed during the model determination have been highlighted. The obtained model is:

\begin{equation} \label{ex_model_restricted}
    z_t = 0.611 \, z_{t-1} + 0.198 \, z_{t-2} + 0.009 \, z_{t-5} + 0.633 \, \varepsilon_t.
\end{equation}

Although the obtained restricted AR model given in \eqref{ex_model_restricted} performs slightly worst than the model given in \eqref{ex_model_l_vector}, it still improves the model obtained with the Y-W approach, Equation \eqref{ex_model_YW}. The key idea of the presented analysis is that selecting appropriate AR model schemes through vector $\textbf{j}$ has the potential to improve the global matching between $\gamma_l^{AR}$ and $\gamma_l^{T}$. To the authors knowledge, this strategy has been considered only to a limited extent in a previous work \cite{Krenk2019}, since in that work an exponential model scheme $\textbf{j}=[1,2, ... , 2^{N-1}]$ was assumed for the presented simulations, which leaves the possibility of optimising the AR scheme unexplored.

Finally, to gather the results obtained with the examples described in Sections \ref{subsec_fromVar2AR} and \ref{subsec_fromVar2AR_restricted}, Figure \ref{fig_errors_summary} shows $|e_l| = |\gamma^T_l - \gamma^{AR}_l|$, computed for the AR model obtained with the Y-W approach, Equation \eqref{ex_model_YW}, the AR model obtained by selecting autocovariance equations $\textbf{l}=[1,2,5]$, see Equation \eqref{ex_model_l_vector}, and the restricted models AR(5,[1,2,5]) obtained with the linear formulation, Equation \eqref{ex_model_restricted}, and the non-linear formulation, Equation \eqref{ex_AR125_0135_nonlinear}.

\begin{figure}[ht!]
    \centering
    \includegraphics[width=0.75\textwidth]{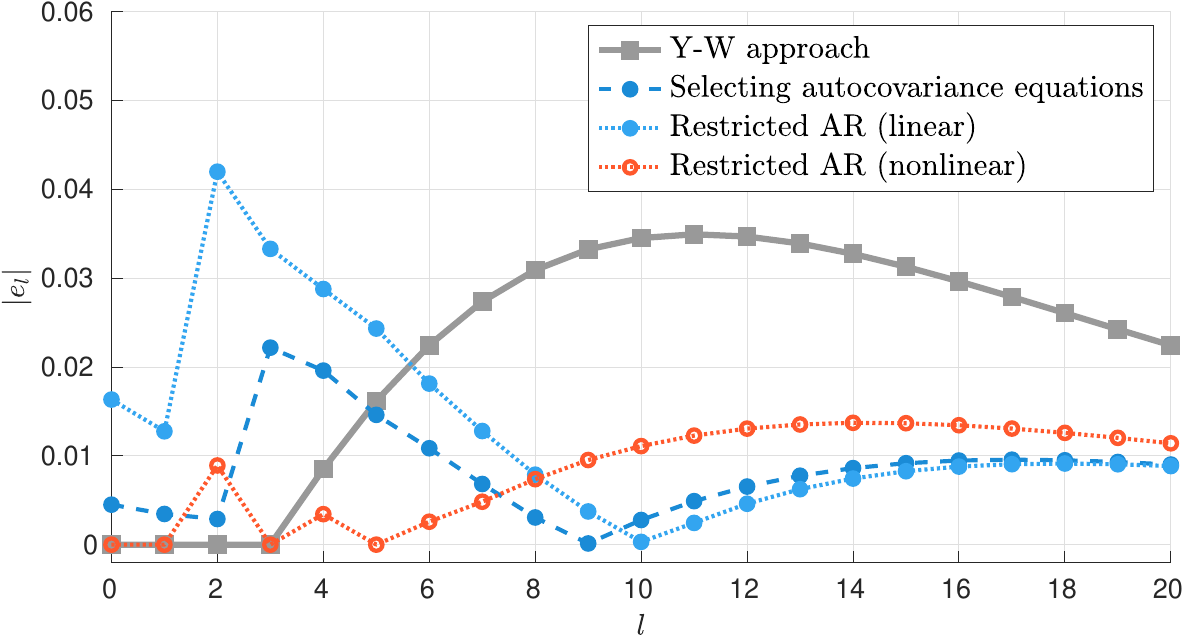} \label{fig_errors_summary}
    \caption{Absolute error between target and theoretical autocovariance function, $|e_l|$, for the AR model obtained with the Y-W approach, Equation \eqref{ex_model_YW}, the AR model obtained by selecting autocovariance equations $\textbf{l}=[1,2,5]$, see Equation \eqref{ex_model_l_vector}, and the restricted models AR(5,\textbf{j}=[1,2,5]) obtained with the linear formulation, Equation \eqref{ex_model_restricted}, and the non-linear formulation, Equation \eqref{ex_AR125_0135_nonlinear}.} \label{fig_errors_summary}
\end{figure}


\section{General formulation for the determination of a restricted AR model from a predefined target autocovariance function} \label{sec_general_formulation} 

This section provides a general formulation for the linear problems addressed in sections \ref{subsec_fromVar2AR} and \ref{subsec_fromVar2AR_restricted}. The non-linear problem described in Section \ref{subsec_fromVar2AR_restricted} is not covered by the following formulation and will be further analysed in future research. 
The objective here is to obtain the parameters of a restricted AR($p,\textbf{j}$) model given a target autocovariance function, from generic input vectors $\textbf{j}=[j_1 , j_2 , ... , j_N]$ (regression coefficients considered in the model) and $\textbf{l}=[l_1 , l_2 , ... , l_N]$ (autocovariance equations considered to obtain the model parameters), by means of a linear system that can be computed with low computational resources.

\

The problem is divided into two steps: 

\begin{itemize}
    \item [(i)] Determination of the regression coefficients $a_{j_i}$, for $i = 1,...,N$, by means of the set of $N$ autocovariance equations for lags gathered in $\textbf{l}$. For convenience, the regression coefficients are encapsulated into a row vector $\textbf{a}=[  a_{j_1} , a_{j_2} , ... , a_{j_N}  ]$. 
    \item [(ii)] Determination of the noise coefficient, $\sigma$, by means of the autocovariance equation for lag $l=0$. 
\end{itemize}

Note that, while in sections \ref{subsec_fromVar2AR} and \ref{subsec_fromVar2AR_restricted} a single system of equations including (i) and (ii) was considered, the division of the problem proposed in this section is always possible since the noise coefficient appears only in the autocovariance equation for lag $l=0$. By doing so, a more handy general formulation is obtained. 
Note also that the formulation is presented for the case of a restricted AR model, but the case of an unrestricted AR model is included by simply considering Equation \eqref{r2u}.

\subsection{Determination of the model coefficients: \textbf{a}} \label{subsub_a}

The autocovariance equations considered for the lags gathered in $\textbf{l}$ and for the case of a restricted AR($p,\textbf{j}$) model can be written in matrix form as follows:

\begin{equation} \label{eq_general_form_a}
    [ ~ \gamma_{l_1} ~ \gamma_{l_2} ~ \dots ~ \gamma_{l_N} ~ ]
    = 
    [ ~ a_{j_1}  ~ a_{j_2}  ~ \dots ~ a_{j_N} ~ ] 
    \begin{pmatrix}
    \gamma_{l_1-j_1} & \gamma_{l_2-j_1}  &  \dots  & \gamma_{l_N-j_1} \\
    \gamma_{l_1-j_2} & \gamma_{l_2-j_2}  &  \dots  & \gamma_{l_N-j_2} \\
    \vdots           & \vdots            &  \ddots & \vdots \\
    \gamma_{l_1-j_N} & \gamma_{l_2-j_N}  &  \dots  & \gamma_{l_N-j_N} \\
    \end{pmatrix},
\end{equation}

\noindent or, in a more compact way:

\begin{equation} \label{eq_general_form_a_matrix}
\bgamma_{\bl} =  \ba \cdot \bgamma_{\bj,\bl} ,
\end{equation}

\noindent where $\bgamma_{\bl}$ is a row vector with dimension $N$, and $\bgamma_{\bj,\bl}$ is a matrix $ N \times N$, both containing values of the autocovariance function at specific time lags, according to vectors \textbf{j} and \textbf{l}. It is worth noting that, for the particular case $\textbf{j}=\textbf{l} = [1, 2, 3, ..., p]$, the system of equations given in \eqref{eq_general_form_a} becomes the Yule-Walker equations.
If the autocovariance terms in \eqref{eq_general_form_a_matrix} are replaced by the target values, the model coefficients are given by:

\begin{equation} \label{eq_general_form_a_matrix_v2}
\ba = \bgamma_{\bl} \cdot \bgamma_{\bj,\bl}^{-1}.
\end{equation}

\subsection{Determination of the noise coefficient: $b$} \label{subsub_b}

The autocovariance equation for time lag $l=0$ for a restricted AR($p,\textbf{j}$) model is as follows:

\begin{equation} \label{eq_general_form_b}
    \gamma_{0}
    = 
    [ ~ a_{j_1}  ~ a_{j_2}  ~ \dots ~ a_{j_N} ~ ] 
	[ \gamma_{-j_1}  \gamma_{-j_2}  \dots   \gamma_{-j_N} ]'
    +
    b^2,
\end{equation}

\noindent where tilde means transposed. Operating and applying symmetry in the autocovariance function, the noise coefficient is given by:

\begin{equation} \label{eq_general_form_b_v2}
    b^2 
    =
    \gamma_0
    -
    [ ~ a_{j_1}  ~ a_{j_2}  ~ \dots ~ a_{j_N} ~ ] 
    \,
    [ \gamma_{j_1} ~ \gamma_{j_2}  ~ \dots ~   \gamma_{j_N} ]' ,
\end{equation}

\noindent or, in a more compact way:

\begin{equation} \label{eq_general_form_b_matrix}
b^2               = \gamma_{0} - \ba \cdot \bgamma_{\bj}',
\end{equation}

\noindent where $\bgamma_{\bj}$ is a row vector containing $N$ values of the target autocovariance function at time lags gathered in $\textbf{j}$.

\

Finally, note that equations \eqref{eq_general_form_a_matrix} and \eqref{eq_general_form_b_matrix} particularised for $\textbf{j}=\textbf{l}=[1,2,...,2^{N-1}]$ represent the univariate case of the formulation introduced in \cite{Krenk2019}. The formulation here presented is more flexible, as it allows searching for optimal \textbf{j} and \textbf{l} vectors independently, that is, without assuming the constraint $\textbf{j}=\textbf{l}$.


\section{Optimal AR models for synthetic isotropic turbulence generation} \label{sec_40_GAs}

In this section, a methodological proposal is presented to obtain optimal AR models from a predefined target autocovariance function. The particular case of homogeneous stationary isotropic (SHI) turbulence is considered. The methodology combines the general formulation described in Section \ref{sec_general_formulation} with the use of genetic algorithms. 
The aim is to find, for a given number of regression coefficients $N$, optimal vectors $\textbf{j} = [ \, j_1 , j_2 , ... , j_N \, ]$ and $\textbf{l} = [ \, l_1 , l_2 , ... , l_N \, ]$.
In this context, \textit{optimal} means that the obtained AR model resulting from expressions \eqref{eq_general_form_a_matrix_v2} and \eqref{eq_general_form_b_matrix} provides minimum mean squared error, $MSE$, between its theoretical autocovariance function, $\gamma^{\text{AR}}_l$, and the target autocovariance function, $\gamma^T_l$. $MSE$ is given by:

\begin{equation} \label{eq_def_MSE}
    MSE =  \frac{1}{M} \sum_{l=0}^{M} e^2_l,
\end{equation}

\noindent with

\begin{equation} \label{eq_def_MSE_e2}
    e_l = \gamma^T_l - \gamma^{\text{AR}}_l.
\end{equation}

$\gamma^T_l$ is defined as the non-dimensional longitudinal autocovariance function of SHI turbulence, $\mathring{R}_u(\mathring{r}) $, as described in Appendix \ref{A_isotropic}; $\mathring{r}=r/L$ is a non-dimensional spatial coordinate, and $L$ is the length scale parameter of the three-dimensional energy spectrum. 
In \eqref{eq_def_MSE}, $M$ represents the maximum lag considered in the computation of the $MSE$. 
This parameter has been fixed to $M=41$, which corresponds to a maximum non-dimensional spatial distance of $ \mathring{r}_{max} = 40 \cdot \Delta \mathring{r} = 4.98 $. It holds that:

$$
\int_0^{\mathring{r}_{max}} \mathring{R}_u(\mathring{r}) \, \text{d} \mathring{r}  \approx 0.995 L_u^x,
$$

\noindent meaning that the selected $M$ value accounts for the $99.5\%$ of the integral length scale, $L_u^x$.

The genetic algorithm is designed to minimise the criterion given in \eqref{eq_def_MSE}. The following inequality constraints were included in the algorithm:

\begin{enumerate}
    \item $0 < j_1 < j_2 < ... < j_N$, with $j_i \in \mathbb{N}$, for $i=1,...,N.$
    \item $0 < l_1 < l_2 < ... < l_N$, with $l_i \in \mathbb{N}$, for $i=1,...,N.$
    \item $j_i - \Delta \le l_i \le j_i + \Delta$, with $\Delta \in \mathbb{N}$ , for $i=1,...,N.$
\end{enumerate}

The first and second constraints derive from the definition of vectors $\textbf{j}$ and $\textbf{l}$. The third constraint introduces the parameter $\Delta$, that regulates the maximum difference between the regression lags included in the AR model and the corresponding autocovariance equations included in the equation system that provides the model parameters. It has been observed that large differences between elements $j_i$ and $l_i$ may derive into numerical instabilities during the optimisation process. Note that $\Delta =0$ means that $\textbf{l} = \textbf{j}$.

The analysis includes a benchmark exercise for a number of models, which are compared for the same model parsimony. 
The range $N=1,...,10$ is considered in what follows. The benchmark comparison includes the following models (italic letters are employed to denote the models):

\begin{itemize}
\item \textit{Y-W}, an unrestricted AR($p$) model obtained through the Yule-Walker approach, that is, $p=N$ and $\textbf{j} = \textbf{l} = [ \, 1 , 2 , ... , p \, ]$.
\item \textit{K-M}, a restricted AR($p,\textbf{j}$) with an exponential model scheme, $\textbf{j} = [ \, 1 , 2 , ... , 2^{N-1} \, ]$), and $\textbf{j} = \textbf{l}$, as proposed in \cite{Krenk2019}.
\item \textit{GA}-0: obtained with a genetic algorithm with $\Delta=0$. This model allows exploring the potential improvement due to relaxing the exponential model scheme employed in \textit{K-M}, while keeping the constraint $\textbf{j} = \textbf{l}$. 
\item \textit{GA}-10: obtained with a genetic algorithm with $\Delta=10$. This model allows assessing the additional improvement with respect to \textit{GA}-0 due to relaxing the constraint $\textbf{j} = \textbf{l}$.
\end{itemize}

\begin{figure}[ht!]
    \centering
    \includegraphics[width=9cm]{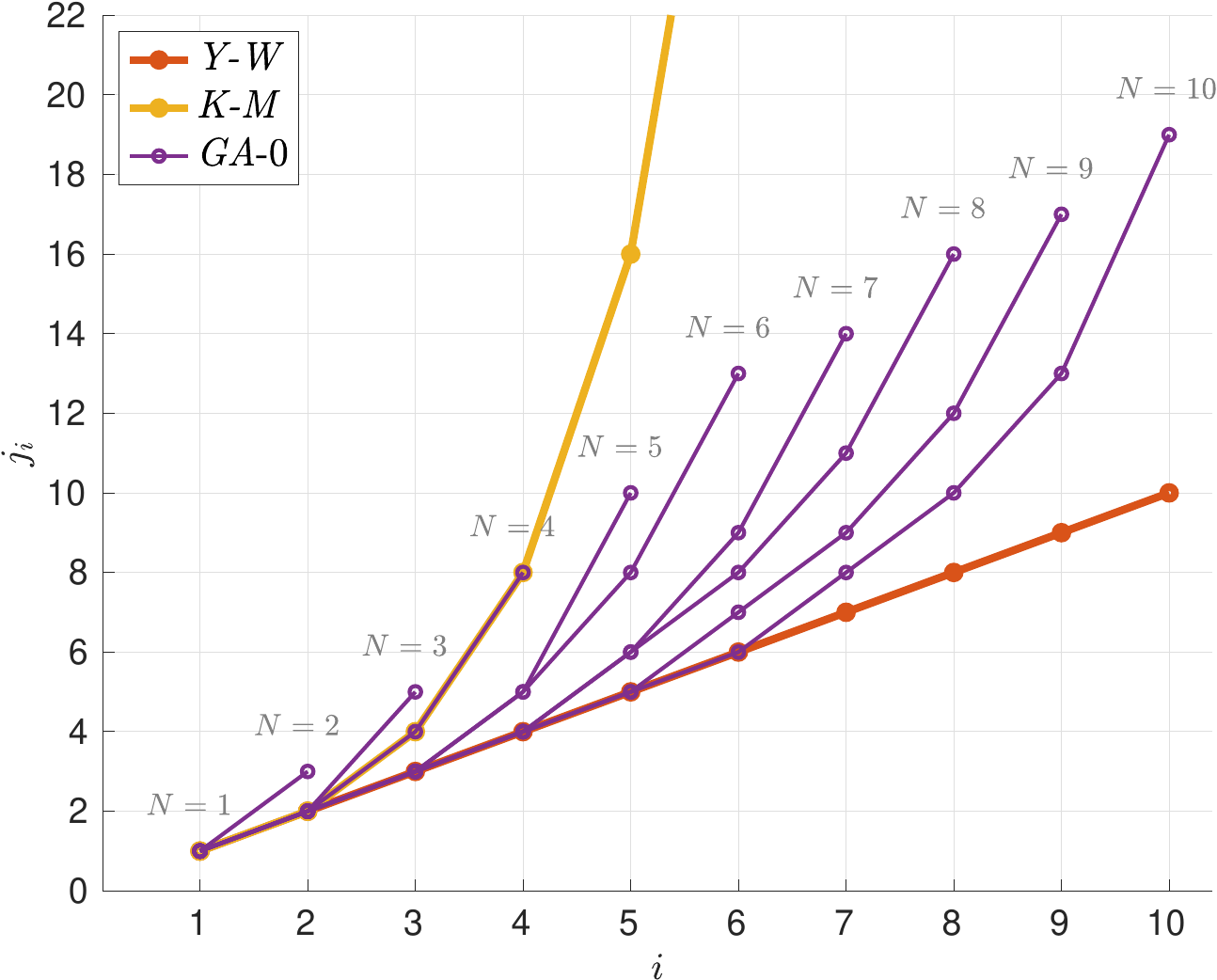}
    \caption{Components $j_i$ of vector $\textbf{j}$ obtained with \textit{GA}-0 model. \textit{Y-W} and \textit{K-M} models are included for comparison.}
    \label{fig_j_vector_GA-0}
\end{figure}

\begin{figure}[ht!]
    \centering
    \includegraphics[width=14cm]{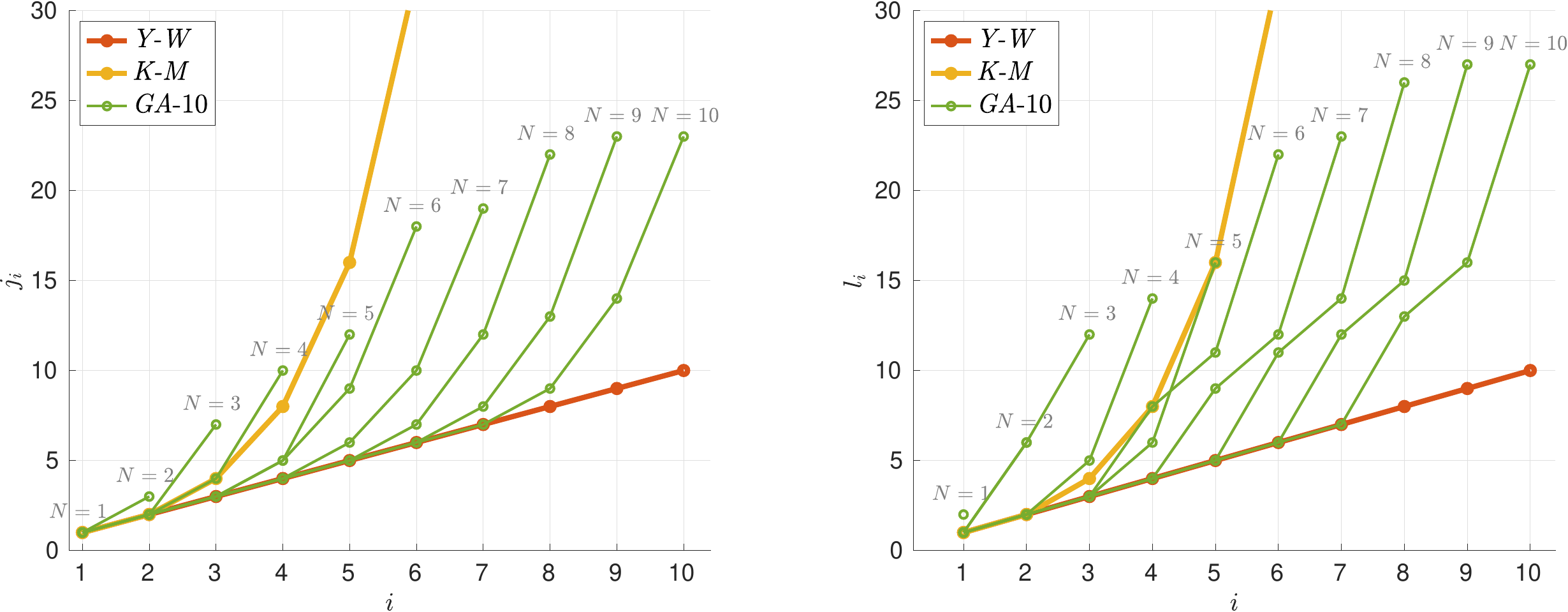}
    \caption{Components $j_i$ of vector $\textbf{j}$ (left) and $l_i$ of vector $\textbf{l}$ (right) obtained with \textit{GA}-10 model. \textit{Y-W} and \textit{K-M} models are included for comparison.}
    \label{fig_j_vector_GA-10}
\end{figure}

Figure \ref{fig_j_vector_GA-0} shows the components of vector $\textbf{j}$ obtained for model \textit{GA}-0, for different $N$ values. 
Figure \ref{fig_j_vector_GA-10} shows the components of vectors $\textbf{j}$ (left) and $\textbf{l}$ (right) obtained for model \textit{GA}-10, for different $N$ values. \textit{Y-W} and \textit{K-M} models are included in both figures for comparison.

It is noted that, for the \textit{GA}-10 model, the maximum difference found between \textit{i}-th elements $j_i$ and $l_i$ was 5, meaning that the choice $\Delta = 10$ was flexible enough to allow the search for optimal AR models witout actually constraining the difference between corresponding elements in \textbf{j} and \textbf{l} vectors.

Results show that, for the most parsimonious models, $N=1$, the single regression term considered in the four models is the previous lag, $\textbf{j}=[1]$. However, for \textit{GA}-10 the employed autocovariance equation is for lag $\textbf{l}=[2]$. 
For the case $N=2$, both \textit{GA}-0 and \textit{GA}-10 models provided optimal regression lags $\textbf{j}=[1, 3]$, different from \textit{Y-W} and \textit{K-M} models, for which $\textbf{j}=[1, 2]$.
Concerning the model order $p$, given by the last term of $\textbf{j}$, $j_N$, results show that, while for models \textit{Y-W} and \textit{K-M} $p$ is determined by $N$ (linearly and exponentially, respectively), the proposed methodology has the ability to reveal an optimal model order for a given model parsimony. It is noted that the obtained optimal order models depend on the specific target autocovariance function considered. Indeed, for increasing $N$ values, the obtained model order for the most flexible model, \textit{GA}-10, stagnates around $p=23$. This value is coherent with the fact that the target autocovariance function is already very close to zero for this lag, see Figure \ref{fig_Annex_VK}. Thus, including additional regression coefficients by reducing the model parsimony is better exploited by the model by rearranging terms in vector $\textbf{j}$ rather than increasing the model order beyond this value, as it is the case for models \textit{Y-W} and \textit{K-M}.

\begin{figure}[h!]
    \centering
    \includegraphics[width=9cm,clip]{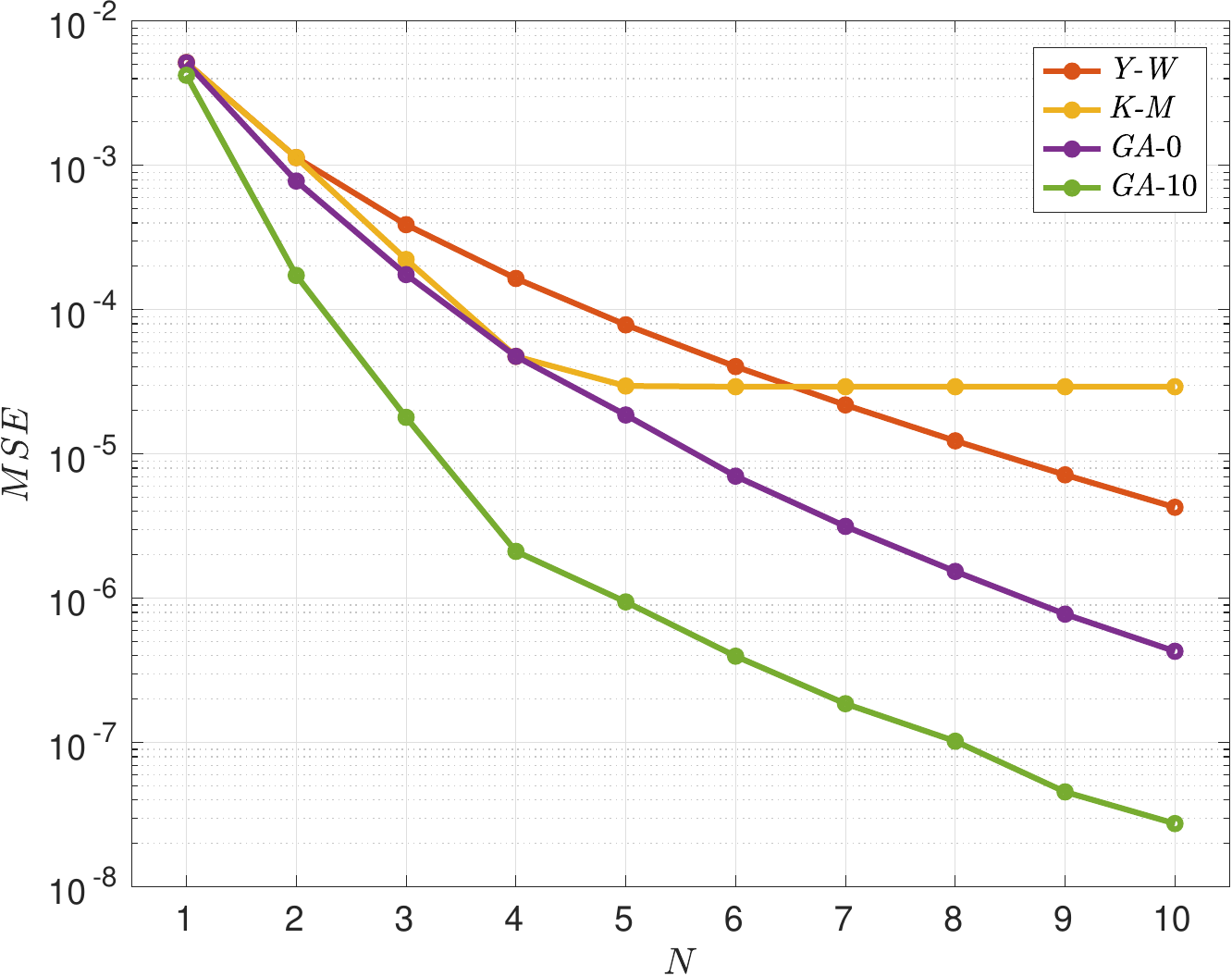}
    \caption{$MSE$ obtained with \textit{Y-W}, \textit{K-M}, \textit{GA}-0 and \textit{GA}-10 models, as a function of the number of regression terms considered in the model, $N$. (Logarithmic scale).}
    \label{fig_S_YW_Krenk_GA-0_GA-10}
\end{figure}

Figure \ref{fig_S_YW_Krenk_GA-0_GA-10} shows the $MSE$ obtained with all the models considered in the analysis, as a function of $N$. 

As expected, the \textit{MSE} obtained with the Y-W approach decreases with $N$, because the obtained models match exactly the target autocovariance function for lags up to the model order. 
For the K-M approach, the error decreases rapidly for low $N$ values, reflecting the advantages of restricted AR models as compared with the Y-W approach. However, the obtained error stagnates from $N=5$ on. This can be explained by the fact that including a regression term for lag $2^{6-1}=32$ (i.e. $N=6$) improves little the fitting of a target autocovariance function that becomes very close to zero for such lag values. This result clearly shows that an exponential model scheme, as that of model \textit{K-M}, has an intrinsic upper limit on the number of regression terms that is worthy to consider, this limit being related to the range of lags for which $\gamma^T_l$ takes non-negligeable values. This range depends on the selected $\Delta \mathring{r}$ employed to obtain the discrete target values from the continuous non-dimensional target autocovariance function of the underlying random process, see Appendix \ref{A_isotropic}.
The improvements provided by \textit{GA}-0 with respect to \textit{K-M} model show the importance of searching for optimal model schemes rather than assuming a predefined model structure. The improvement becomes noticeable for $N$ values larger than the aforementioned upper limit $N=5$.  
Finally, the gap between the results for \textit{GA}-0 and \textit{GA}-10 clearly shows the additional improvement associated with an optimal selection of the autocovariance equations employed in the determination of the model parameters. Thus, relaxing the hypothesis $\textbf{l} = \textbf{j}$, that usually underlies in the literature, is actually a clear path for improvement.

\begin{figure}[ht!]
    \centering
    \includegraphics[width=9cm]{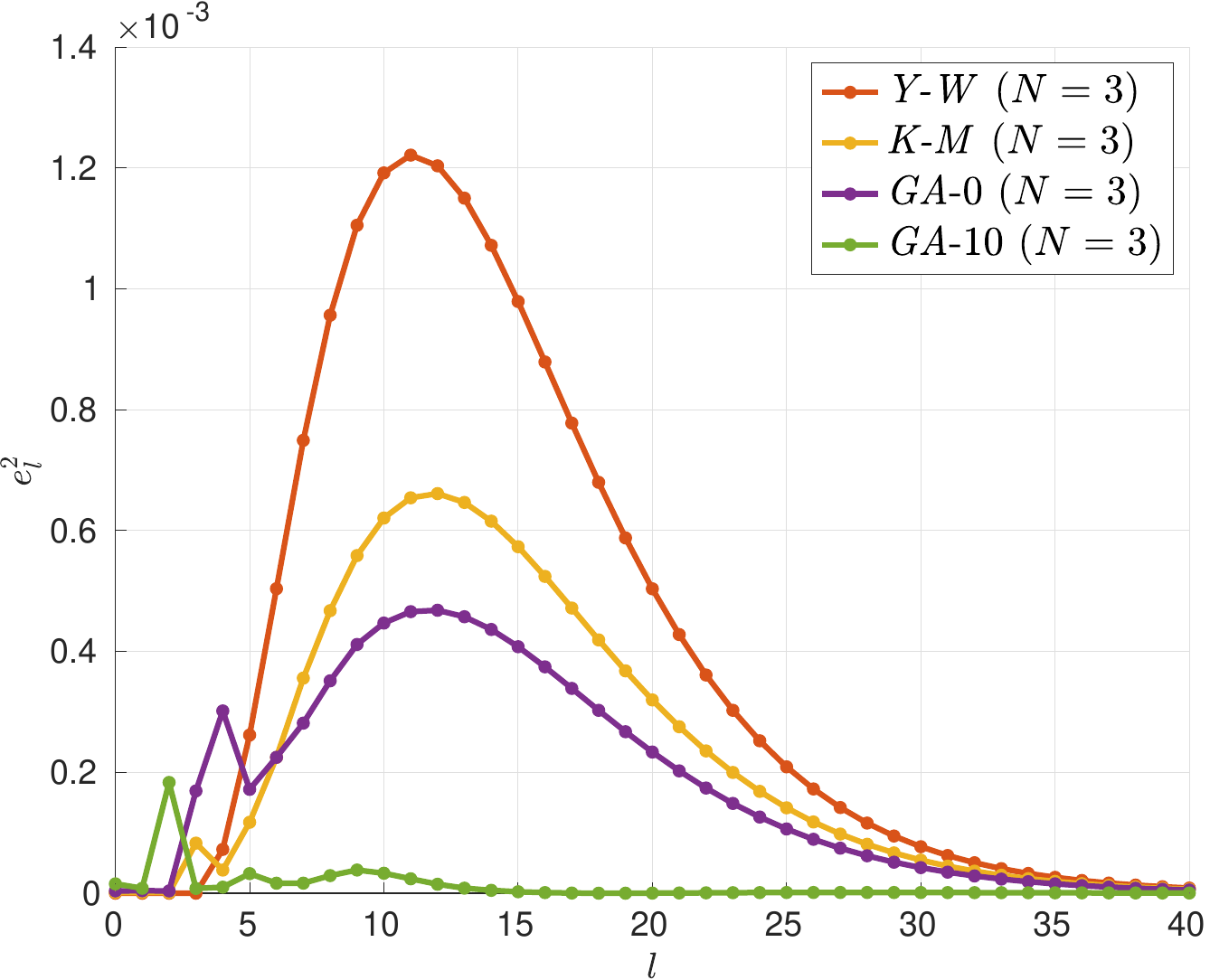}
    \caption{Squared fitting errors $e^2(l)$ for the different models. Case $N=3$.}
    \label{fig_distributed_errors_YW_Krenk_GA_0_GA_10_N_3_lambda_30}
\end{figure}

To delve more into this analysis, the distribution of $MSE$ with the lag is analysed. Figure \ref{fig_distributed_errors_YW_Krenk_GA_0_GA_10_N_3_lambda_30} shows $e^2_l$, see Equation \eqref{eq_def_MSE_e2}, for case $N=3$.

\begin{figure}[ht!]
    \centering
    \includegraphics[width=14cm]{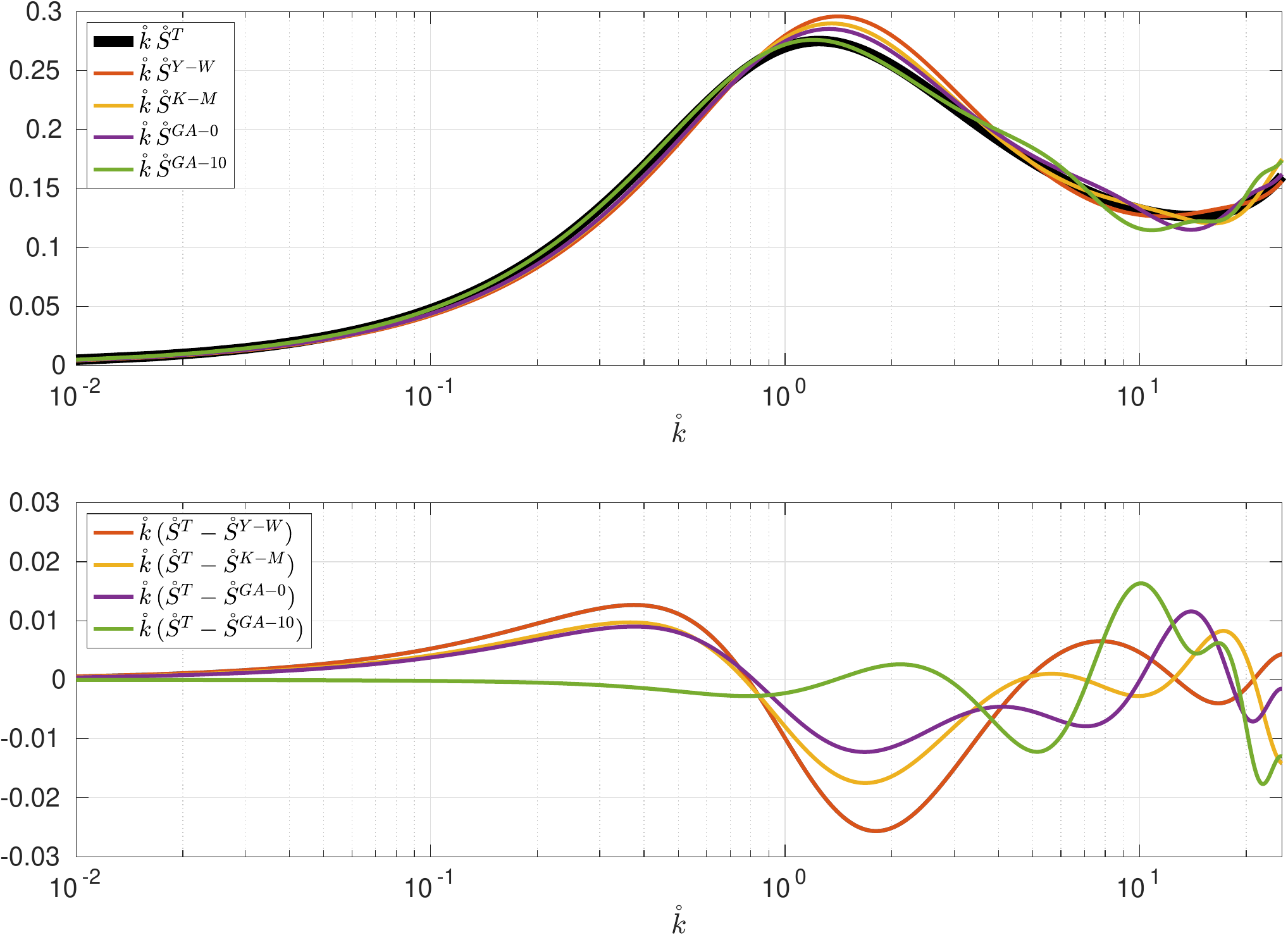}
    \caption{Top: non-dimensional one-sided target spectrum, $\mathring{S}^{T}(\mathring{k})$, together with the corresponding theoretical spectra of the AR models obtained with the four approaches; bottom: differences between the target spectrum and the AR model spectra. Case $N=3$.}
    \label{fig_spectrum_YW_Krenk_GA_0_GA_10_N_3_lambda_30}
\end{figure}

It can be seen that improving the global fitting of the target autocovariance function by means of restricted AR models may come at the expense of deteriorating the local fitting in some of the first lags, as compared with the results of the unrestricted models employed under the Y-W approach. For example, model \textit{GA}-10 improves greatly the other models in terms of the global fitting, see Figure \ref{fig_S_YW_Krenk_GA-0_GA-10}, but at the same time \textit{GA}-10 shows the highest error for lag $l=2$. This trade-off between global fitting and local fitting of the target autocovariance function is relevant when using restricted AR models, and it probably needs to be addressed with a broader concept of \textit{optimal} model, which may vary from one problem to another. 
In this regard, additional information from the frequency domain could be included to define the notion of optimal model. This is particularly relevant for some engineering problems, for which some constraints could be defined in terms of frequency ranges, like those involving wind loads \cite{Li1990,Kareem2008}.
For illustrative purposes, Table \ref{table_comparison_models_N_3} shows the model parameters obtained for $N=3$, and Figure \ref{fig_spectrum_YW_Krenk_GA_0_GA_10_N_3_lambda_30} gathers, on top, the non-dimensional one-sided target spectrum, $\mathring{S}^{T}(\mathring{k})$, together with the corresponding theoretical spectra of the AR models obtained with the four approaches, and at the bottom, the differences between the target spectrum and the AR model spectra.  $\mathring{k}$ is the non-dimensional wave number, $\mathring{k}= k \, L$. Note that $\mathring{S}^{T}(\mathring{k})$ is affected by aliasing due to the discretisation of the target autocovariance function, see details in Appendix \ref{A_isotropic}. The non-dimensional one-sided spectrum of an AR model can be obtained either from the model coefficients,

\begin{equation}
\mathring{S}^{AR} (\mathring{k}) = \frac{1}{\mathring{k}_{max}} \frac{ b^2 / \sigma_0^2 }{ \left\lvert 1 - \sum_{n=1}^N a_{j_n} \, \text{exp} \left( -i\,n \, \pi \frac{\mathring{k}}{\mathring{k}_{max}} \right) \right\rvert^2 },
\end{equation}

\noindent or from its theoretical autocovariance function \cite{Box2016}:

\begin{equation}
\mathring{S}^{AR} (\mathring{k}) = \frac{1}{\mathring{k}_{max}} \left[  \gamma_0^{AR} + 2 \sum_{l=1}^{\infty} \gamma^{AR}_l \, \text{cos} \left( \frac{\pi l\,\mathring{k}}{\mathring{k}_{max}} \right) \right] ,
\end{equation}

\noindent where $\mathring{k}_{max} = \frac{1}{2 \Delta \mathring{r}}$ is the maximum non-dimensional wave number, and $\Delta \mathring{r}$ is the non-dimensional length employed to obtain the discrete version of the target autocovariance function, see Appendix \ref{A_isotropic}.

\begin{table}[h]
\centering
 \caption{AR models obtained for the different approaches for $N=3$.} \label{table_comparison_models_N_3}
 \begin{tabular}{l l l c} 
 \hline
 Approach   & $~~~\textbf{j}$    & $~~~\textbf{l}$    & Model \\  
 \hline
\textit{Y-W}   & [1,2,3] & [1,2,3]  & $z_t = 0.663 \, z_{t-1} + 0.099 z_{t-2} + 0.044 \, z_{t-3} + 0.636 \, \varepsilon_t $\\
\textit{K-M}   & [1,2,4] & [1,2,4]  & $z_t = 0.664 \, z_{t-1} + 0.109 z_{t-2} + 0.035 \, z_{t-4} + 0.636 \, \varepsilon_t $\\ \textit{GA}-0  & [1,2,5] & [1,2,5]  & $z_t = 0.665 \, z_{t-1} + 0.115 z_{t-2} + 0.029 \, z_{t-5} + 0.636 \, \varepsilon_t $\\ \textit{GA}-10 & [1,2,7] & [1,6,12] & $z_t = 0.646 \, z_{t-1} + 0.147 z_{t-2} + 0.025 \, z_{t-7} + 0.635 \, \varepsilon_t $\\ \hline
 \end{tabular}
\end{table}

Figure \ref{fig_spectrum_YW_Krenk_GA_0_GA_10_N_3_lambda_30} illustrates that, in terms of fitting the target spectrum, model \textit{GA}-10 outperforms the rest of the models for low frequencies (up to $\mathring{k} = 4$), including the frequency at which the maximum of the target spectrum is attained ($\mathring{k} \approx 1.245$). For higher frequencies, all model spectra show similar oscillations around the target spectrum. This result suggests that, a priory, reducing the local fitting of the target autocovariance function in the first lags to improve the global fitting does not have a clear negative impact on the spectrum fitting.


\section{Generalisation to multivariate processes} \label{sec_40_VARs}

In this section, the analyses described in sections \ref{sec_AR}, \ref{sec_AR_restricted} and \ref{sec_general_formulation} are extended for a one-dimensional multivariate random process, $\{\bz_t(\alpha)\}$, with the column vector $\bz_t = [z_1 , z_2 , ... , z_k ]_t'$ being a $k$-variate random variable. In particular, we describe the reasons why some bottlenecks in terms of computational cost may arise as a consequence of the multivariate character of the formulation. Note that multi-dimensional multivariate problems in some cases admit a one-dimensional multivariate formulation, as firstly described in \cite{MignoletSpanos1991_I}. 
In \cite{Krenk2019}, a three-dimensional three-variate description of SHI turbulent wind field was expressed in the form of a one-dimensional $3P$-variate random process by stacking the three velocity components at $P$ points of the plane perpendicular to the mean wind into a random variable.

The general formulation of a zero-mean $k$-variate VAR($p$) model is:

\begin{equation}\label{eq_VAR}
    \bz_t =  \sum_{i=1}^p \bPhi_i \, \bz_{t-i} +  \bSigma \, \bEpsilon_t .
\end{equation}

\noindent where  $\bz = [z_1 , z_2 , ... , z_k ]'$ is a $k$-variate random variable, and $\bEpsilon_t$ is a sequence of independent and identically distributed (iid) random vectors with zero mean, $\E[\bEpsilon_t] = \bzero_{k\times1}$, and unity covariance matrix, $\Var[\bEpsilon_t]=\bI_k$. The $p+1$ model matrix parameters are given by the regression matrices, $\bPhi_i$ with $i=1,...,p$, and the noise matrix, $\bSigma$; the dimension of all the matrix parameters is $k \times k$. The covariance of the random term is given by $\text{Var}[\bSigma \, \bEpsilon_t ] = \E[ (\bSigma \, \bEpsilon_t) \, (\bSigma \, \bEpsilon_t)' ] = \bSigma \bSigma'$.

$\bGamma_{t_1,t_2}$ is the covariance between random vectors $\bz_{t_1}$ and $\bz_{t_2}$. As in the univariate case, the assumption of stationary process implies that $\bGamma_{t_1,t_2}$ depends only on the time lag, $l=t_1-t_2$, which is referred to as the covariance matrix function:

\begin{equation} \label{eq_covariance_definition}
     \bGamma_l = \Cov [ \, \bz_t , \bz_{t-l} \, ] = \E[ \, \bz_t \, \bz_{t-l}' \, ] = 
     \begin{pmatrix}
     \gamma_{z_1,z_1,l} & \gamma_{z_1,z_2,l} & \dots  & \gamma_{z_1,z_k,l}   \\
     \vdots              & \vdots              & \ddots        & \vdots                \\
     \gamma_{z_k,z_1,l} & \gamma_{z_k,z_2,l} & \dots & \gamma_{z_k,z_k,l}    \\
     \end{pmatrix}.
\end{equation}



The covariance equations that generalise equations \eqref{eq_AR_variance_0} and \eqref{eq_AR_variances} for the multivariate case are:

\begin{align} \label{eq_VAR_cov0}
    \bGamma_0 & = \sum_{i=1}^p \bPhi_i \, \bGamma_{-i} + \bSigma \bSigma',
\end{align}

\begin{align} \label{eq_VAR_covariances}
    \bGamma_l = \sum_{i=1}^p \bPhi_i \, \bGamma_{l-i}.   
\end{align}

\subsection{Computing the theoretical covariance matrix function of a VAR model}

As in the univariate case, equations \eqref{eq_VAR_cov0} and \eqref{eq_VAR_covariances} can be employed to compute the theoretical covariance matrix function, $\bGamma^{VAR}_l$, from the model parameters, $\bPhi_i, i=1,...,p$ and $\bSigma$. However, note that in the multivariate case, $\bGamma_{-l} = \bGamma_l'$. This fact makes the linear formulation described in Section \ref{subsec_fromAR2Var} not applicable for the multivariate case, as it is not possible to replicate the step given between equations \eqref{eq_fromAR2Vars} and \eqref{eq_fromAR2Vars_v0}. There are two approaches for obtaining the theoretical covariance matrix function of a VAR($p$) model, both involving an increasing computational cost, as compared with the univariate case:

\begin{enumerate}
    \item To obtain $\bGamma^{VAR}_l$ from the covariance matrix of the VAR(1) representation of the VAR($p$) model. This strategy provides the exact covariance matrix function of the VAR($p$) model, but it is computationally very expensive, as the size of the involved matrices scales up to $(pk)^2 \times (pk)^2$.
    \item To obtain $\bGamma^{VAR}_l$ from the covariance matrix function of the equivalent multivariate Vector Moving Average (VMA) model. Since this equivalent VMA model contains infinite terms, truncation to a VMA($q$) model is required, meaning that the obtained covariance matrix function is an approximation of the exact VAR covariance matrix function. In this case, the increase in the computational cost comes from the fact that appropriate $q$ values are related to the integral length scale of the process, which typically leads to $q \gg p$.
\end{enumerate}

In what follows, both methodologies are briefly described.

\subsubsection{Covariance matrix function of a VAR($p$) model through the VAR(1) representation} \label{par_VAR1}

A $k$-variate VAR($p$) model can be expressed in the form of an extended $pk$-dimensional VAR(1) model by using an expanded model representation \cite{Tsay2013}. For illustrative purposes, the case of a VAR(3) model is analysed:

\begin{equation}\label{eq_VAR3}
    \bz_t =   \bPhi_1 \, \bz_{t-1} +  \bPhi_2 \, \bz_{t-2} +  \bPhi_3 \, \bz_{t-3} +  \bSigma \, \bEpsilon_t .
\end{equation}

Let us define the expanded $3k$-variate random variable as:

\begin{equation}
     \bz_t^* =  \begin{pmatrix}  
            \bz_t \\  
            \bz_{t-1} \\
            \bz_{t-2}
            \end{pmatrix}. 
\end{equation}

The VAR(1) representation of the original VAR(3) model is given by:

\begin{equation}\label{eq_VAR1}
    \bz_t^* = \bPhi_1^* \, \bz_{t-1}^*  + \bSigma^* \, \bEpsilon_t^* ,
\end{equation}

\noindent where $\bPhi_1^*$ is called the companion matrix, defined as:

\begin{equation}
    \bPhi_1^* = \begin{pmatrix}
                \bPhi_1  &  \bPhi_2  &  \bPhi_3 \\
                \bI_k      &  \bzero_k   &  \bzero_k \\
                \bzero_k   & \bI_k       &  \bzero_k \\
                \end{pmatrix},
\end{equation}

\noindent and the random matrix is given by: 

\begin{equation}
    \bSigma^* = \begin{pmatrix}
                \bSigma    &  \bzero_k   &  \bzero_k \\
                \bzero_k   &  \bzero_k   &  \bzero_k \\
                \bzero_k   &  \bzero_k   &  \bzero_k \\
                \end{pmatrix} ~,~ 
                    \bEpsilon_t^* =   \begin{pmatrix} \bEpsilon_t \\ \bzero_{k\times1} \\ \bzero_{k\times1} \end{pmatrix}.
\end{equation}

The covariance matrix function of the extended VAR(1) model of equation \eqref{eq_VAR1} is given by:

\begin{align} \label{cov0_VAR1}
    \bGamma^*_0 & = \Var [ \, \bz_t^* , \bz_{t}^* \, ] = \E[ \, \bz_t^* \, {\bz_{t}^*}' \, ] \nonumber \\
                 & =\E[ \, (\bPhi_1^* \, \bz_{t-1}^* + \bSigma^* \, \bEpsilon_t^* ) \, (\bPhi_1^* \, \bz_{t-1}^* + \bSigma^* \, \bEpsilon_t^* )' \, ] \nonumber \\ 
                 & = \bPhi_1^* \, \bGamma^*_0 \, {\bPhi_1^*}' + \bSigma^* {\bSigma^*}'. 
\end{align}

The solution of Equation \eqref{cov0_VAR1} is given by \cite{Tsay2013}:

\begin{equation} \label{eq_cov_VAR1}
    \text{vec}(\bGamma^*_0) = \text{vec}(\bSigma^* {\bSigma^*}') \, (\bI_{(pk)^2} - \bPhi_1^* \otimes \bPhi_1^*)^{-1}, 
\end{equation}

\noindent where vec($\cdot$) denotes the column-stacking vector of a matrix and $\otimes$ is the Kronecker product of two matrices.

The covariance matrix function of the original VAR(3) model can be readily obtained from $\bGamma^*_0$, noting that $\bGamma_0^{VAR}$, $\bGamma_1^{VAR}$ and $\bGamma_2^{VAR}$ are given by:

\begin{equation}
    \bGamma^*_0 = \E[ \, \bz_t^* \, {\bz_{t}^*}' \, ] =    \begin{pmatrix}
                                                       \bGamma_0^{VAR}   &    \bGamma_1^{VAR}   &  \bGamma_2^{VAR}  \\
                                                       \bGamma_{-1}^{VAR}  &    \bGamma_0^{VAR}   &  \bGamma_1^{VAR}  \\
                                                       \bGamma_{-2}^{VAR}  &    \bGamma_{-1}^{VAR}  &  \bGamma_0^{VAR}  \\
                                                       \end{pmatrix},
\end{equation}

\noindent and using recursively Equation (\ref{eq_VAR_covariances}) to obtain $\bGamma^{VAR}_l$ for lags $l \ge 3$.

\subsubsection{Covariance matrix function of a VAR($p$) model through VMA($q$) representation} \label{par_VMAq} 

It can be demonstrated that any VAR($p$) model admits a Vector Moving-Average (VMA) representation with infinite terms:

\begin{equation}\label{eq_VMA}
    \bz_t = \bSigma \, \bEpsilon_t + \sum_{i=1}^{\infty} \bPsi_i \, ( \bSigma \, \bEpsilon_{t-i}),
\end{equation}

\noindent where matrices $\bPsi_i$ are known functions of the VAR regression matrices, $\bPhi_i$ \cite{Tsay2013}:

\begin{equation}\label{eq_VMA_coeffs_from_VAR}
    \bPsi_i = \sum_{j=1}^{\text{min}(i,p)} \bPhi_j \, \bPsi_{i-j} ~ , ~ \text{for} ~ i=1,2,...
\end{equation}

\noindent and $\bPsi_0 = \bI_k$.

The theoretical covariance matrix function of a VMA process is given by:

\begin{align} \label{eq_cov_VMA}
    \bGamma^{VMA(\infty)}_l & = \Cov [ \, \bz_t , \bz_{t-l} \, ] =  \E[ \, \bz_t \, \bz_{t-l}' \, ]  \nonumber \\
    & = \E [ \, ( \bSigma \, \bEpsilon_t + \sum_{i=1}^{\infty} \bPsi_i \, ( \bSigma \, \bEpsilon_{t-i}) ) \, ( \bSigma \, \bEpsilon_{t-l} + \sum_{i=1}^{\infty} \bPsi_i \, ( \bSigma \, \bEpsilon_{t-l-i})  )'  ]   \nonumber \\
    & = \sum_{i=l}^\infty \bPsi_i \, \bSigma \bSigma' \, \bPsi_{i-l}'.
\end{align}

$\bGamma^{VMA(\infty)}_l$ matches exactly the covariance matrix function of the related VAR($p$) model. However, for practical reasons, this approach requires truncation by considering only the first $q$ terms of the VMA representation, that is, a VMA($q$) model:    
    
\begin{equation} \label{eq_VMA_q}
    \bz_t = \bSigma \, \bEpsilon_t + \sum_{i=1}^{q} \bPsi_i \, (\bSigma \, \bEpsilon_{t-i}).
\end{equation}

The covariance matrix function of a VMA($q$) model is given by:

\begin{equation} \label{eq_cov_VMA_q}
    \bGamma^{VMA(\textit{q})}_l  = \Cov[ \, \bz_t , \bz_{t-l} ]  = 
                \begin{cases}
                \sum\limits_{i=l}^q \bPsi_i \bSigma \bSigma' \bPsi_{i-l}' & \text{ , for } 0\le l \le q  \\
                \bzero_k                                     & \text{ , for }  l > q . 
                \end{cases}
\end{equation}

In summary, $\bGamma^{VAR}_l$ can be approximated in a two step procedure: (i) computing a truncated VMA($q$) representation of the VAR$(p)$ model, see Equation \eqref{eq_VMA_q}; the $q$ coefficient matrices of the VMA representation can be computed through Equation \eqref{eq_VMA_coeffs_from_VAR}; (ii) computing the covariance matrix function of the VMA($q$) model given by Equation \eqref{eq_cov_VMA_q}. Concerning the choice for parameter $q$, note that $\bGamma^{VMA(\textit{q})}_l = \bzero_k$ for $l>q$, thus $q$ should be such that $\bGamma^{VAR}_q \approx \bzero_k $.

\begin{figure}[ht!]
    \centering
    \includegraphics[width=0.75\textwidth]{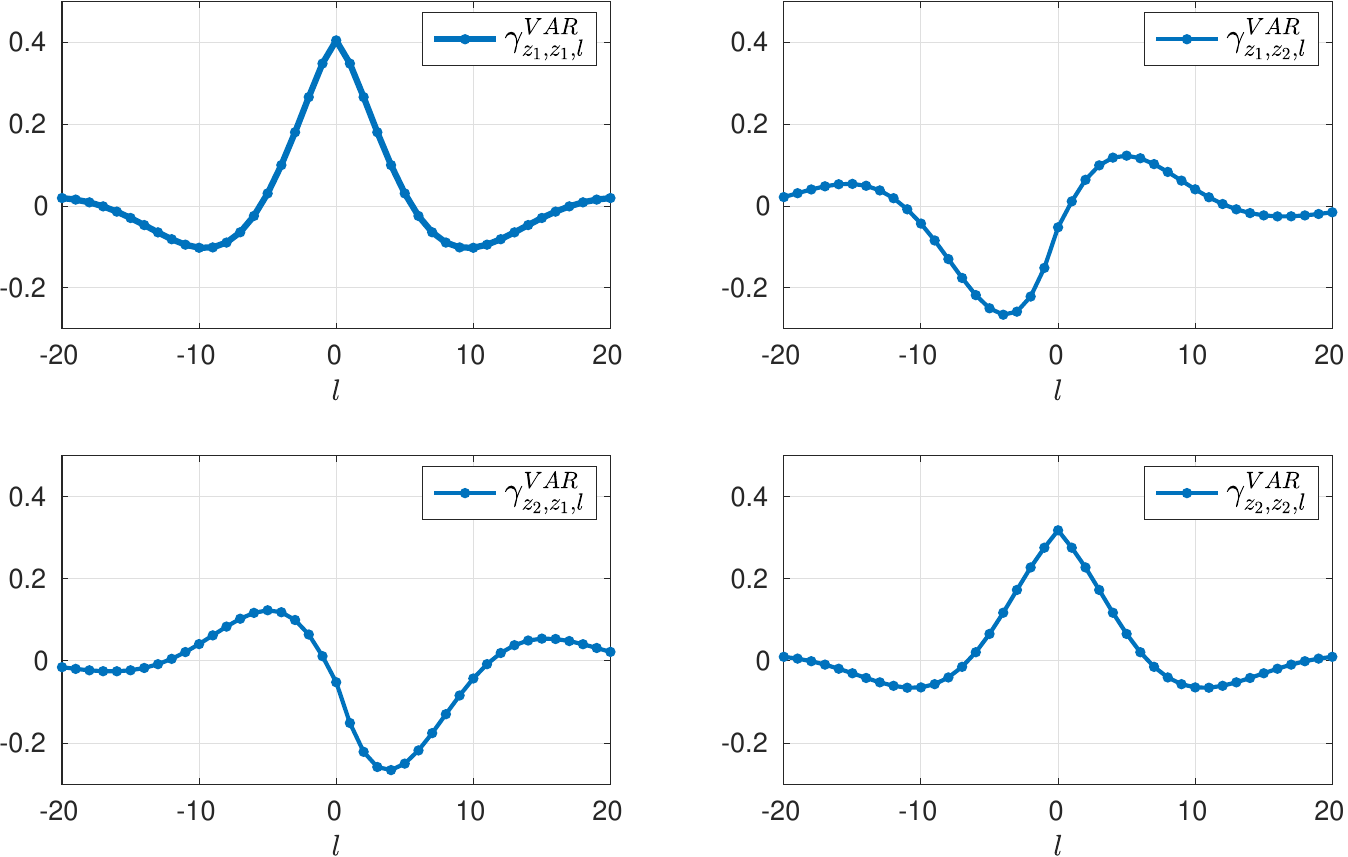}
    \caption{Exact covariance matrix function of a VAR(2) process, obtained through VAR(1) representation.}
    \label{fig_fromVAR2Covars_1}
\end{figure}

\begin{figure}[ht!]
    \centering
    \includegraphics[width=0.75\textwidth]{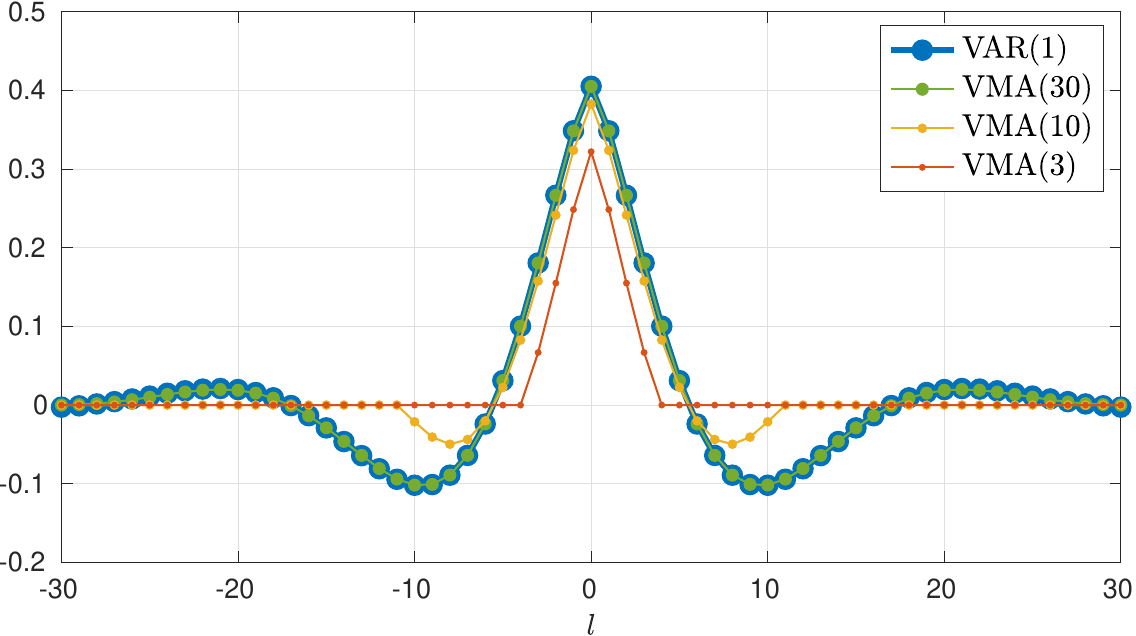}
    \caption{$\gamma^{VAR}_{z_1,z_1,l}$ from the covariance matrix function of a VAR(2) process obtained through VAR(1) representation, and three estimations obtained through the VMA($q$) representation.}
    \label{fig_fromVAR2Covars_2}
\end{figure}

To illustrate the two presented methodologies, figures \ref{fig_fromVAR2Covars_1} and \ref{fig_fromVAR2Covars_2} show the theoretical covariance matrix function, $\bGamma_l^{VAR} $ of the following 2-variate VAR(2) model:

$$ \bz_t =  \begin{pmatrix}
            z_{1,t} \\
            z_{2,t}
            \end{pmatrix}
            =
            \begin{pmatrix}
            1.1   &  -0.1 \\
            -0.2  &   0.7   
            \end{pmatrix}
            \bz_{t-1} +
            \begin{pmatrix}
            -0.3  &  0.2 \\
            -0.1 &   0.1   
            \end{pmatrix}
            \bz_{t-2} 
            + 
            \begin{pmatrix}
            0.3  &  0.0 \\
            0.1 &   0.2   
            \end{pmatrix}
            \bEpsilon_t 
            .$$

\noindent In particular, Figure \ref{fig_fromVAR2Covars_1} shows the exact covariance matrix function computed through the VAR(1) representation, where the property $\bGamma_{-l}=\bGamma_l'$ can be appraised from $\gamma^{VAR}_{z_1,z_2,l}$ and $\gamma^{VAR}_{z_2,z_1,l}$.

Figure \ref{fig_fromVAR2Covars_2} illustrates $\gamma^{VAR}_{z_1,z_1,l}$ together with three approximations obtained through the VMA($q$) representation. It can be noted that a reasonable approximation requires $q \gg p$.

\subsection{General formulation for the determination of a restricted VAR model from a predefined target covariance matrix function}

A restricted VAR($p,\textbf{j}$) model, with $\textbf{j}=[j_1, j_2, ..., j_N]$ is given by:

\begin{equation}\label{eq_VAR_restricted}
    \bz_t = \sum_{i=1}^N  \bA_{j_i} \, \bz_{t-j_i} + \bB \, \bEpsilon_t.
\end{equation}

As in the univariate case, a restricted VAR($p$) model defined by vector $\textbf{j}$ can be seen as a particular case of a VAR($p$) model for which:

$$ \bPhi_h = \begin{cases}
 \bA_h \, , \text{~for~} h \in \textbf{j} \\
 \bzero_k   ~~ , \text{~for~} h \notin \textbf{j} \\
 \end{cases} ,~ h=1,...,p.
$$

and 

$$ \bSigma = \bB .$$

Note also that a VAR($p$) model can be seen as a particular case of a restricted VAR$(p)$ model for which: 

$$ \textbf{j}= [ \, 1, 2, 3, ..., p \, ] .$$

In the following, the formulation for obtaining the matrix parameters of a restricted VAR($p$) model from a target covariance matrix function is introduced for generic vectors $\textbf{j}=[j_1 , j_2 , ... , j_N]$ and $\textbf{l}=[l_1 , l_2 , ... , l_N]$. 
As in the univariate case, the problem is divided into two steps, and the regression matrix coefficients are encapsulated into matrix $\bA=[\bA_{j_1} \, \bA_{j_2} \, ... \, \bA_{j_N}]$, which has dimension $ k \times kN$.

\subsubsection{Determination of the model matrix coefficients: \textbf{A}} \label{subsub_Amatrix}

The multivariate form of Equation \eqref{eq_general_form_a} is given by:

\begin{equation} \label{eq_general_form_A}
    [ ~ \bGamma_{l_1} ~ \bGamma_{l_2} ~ \dots ~ \bGamma_{l_N} ~ ]
    = 
    [ ~ \bA_{j_1}  ~ \bA_{j_2}  ~ \dots ~ \bA_{j_N} ~ ] 
    \begin{pmatrix}
    \bGamma_{l_1-j_1} & \bGamma_{l_2-j_1}  &  \dots  & \bGamma_{l_N-j_1} \\
    \bGamma_{l_1-j_2} & \bGamma_{l_2-j_2}  &  \dots  & \bGamma_{l_N-j_2} \\
    \vdots           & \vdots            &  \ddots & \vdots \\
    \bGamma_{l_1-j_N} & \bGamma_{l_2-j_N}  &  \dots  & \bGamma_{l_N-j_N} \\
    \end{pmatrix},
\end{equation}

\noindent or in a more compact way:

\begin{equation} \label{eq_general_form_A_matrix}
\bGamma_{\bl} = \bA \cdot \bGamma_{\bj,\bl},
\end{equation}

\noindent where $\bGamma_{\bl}$ is a matrix $ k \times kN$ and $\bGamma_{\bj,\bl}$ a matrix $ kN\times kN$. By replacing the covariance terms by the target values, the model matrix coefficients are given by:

\begin{equation}
\bA = \bGamma_{\bl} \cdot \bGamma_{\bj,\bl}^{-1}.
\end{equation}

\subsubsection{Determination of the noise matrix: \textbf{B}} \label{subsub_Bvector}

The multivariate version of Equation \eqref{eq_general_form_b} is:

\begin{equation} \label{eq_general_form_B}
    \bGamma_0
    = 
    [ ~ \bA_{j_1}  ~ \bA_{j_2}  ~ \dots ~ \bA_{j_N} ~ ] 
    [\bGamma_{-j_1} \, \bGamma_{-j_2}  \, \dots  \,  \bGamma_{-j_N} ]'
    +
    \bB\bB'.
\end{equation}

Operating and applying $\bGamma_{-l} = \bGamma_l'$, the covariance matrix of the random term, $\bB\bB'$, is given by:

\begin{equation} \label{eq_general_form_B_v2}
    \bB\bB'
    =
    \bGamma_0
    -
    [ ~ \bA_{j_1}  ~ \bA_{j_2}  ~ \dots ~ \bA_{j_N} ~ ] 
    \,
    [ \bGamma_{j_1} ~ \bGamma_{j_2}  ~ \dots ~   \bGamma_{j_N} ]',
\end{equation}

\noindent or in a more compact way:

\begin{equation} \label{eq_general_form_B_matrix}
\bB\bB' = \bGamma_0 - \bA \cdot \bGamma_{\bj}',
\end{equation}

\noindent where $\bGamma_{\bj}$ is a matrix $k \times kN$. From \eqref{eq_general_form_B_matrix}, the noise matrix $\bB$ can be obtained in several ways, for example, through Cholesky decomposition.
Finally, it is remarked that equations \eqref{eq_general_form_A_matrix} and \eqref{eq_general_form_B_matrix} particularised for \textbf{j}=\textbf{l} represent the formulation introduced in \cite{Krenk2019} (in particular $\bGamma_0 = C_{uu}$, $\bGamma_{\bj}=\bGamma_{\bl}=C_{uw}$ and $\bGamma_{\bj,\bl}=C_{ww}$).

As an example, two VAR models with three regression matrices have been computed to reproduce the target covariance matrix function described in Appendix \ref{A_isotropic}, for the case of the longitudinal wind component at two spatial locations with a non-dimensional lateral separation of $  \Delta \mathring y = \frac{\Delta y}{L} =  \lambda = 0.747$. The first model is a VAR(3) model, and it was computed with the Y-W approach, i.e. $\textbf{j}=\textbf{l}=[1, 2, 3]$:

\begin{equation}
 \bz_t =  
            \begin{pmatrix}
            0.659   &  0.022 \\
            0.022  &   0.659   
            \end{pmatrix}
            \bz_{t-1} +
            \begin{pmatrix}
            0.096  &  0.011 \\
            0.011 &   0.096   
            \end{pmatrix}
            \bz_{t-2} 
	    \begin{pmatrix}
            0.039  &  0.015 \\
            0.015 &   0.039   
            \end{pmatrix}
            \bz_{t-3} 
            + 
            \begin{pmatrix}
            0.634  &  0.000 \\
            0.013 &   0.634   
            \end{pmatrix}
            \bEpsilon_t 
            .
\end{equation}

The second model is a restricted VAR(5,[1,2,5]) computed for \textbf{l}=[1, 2, 6]:

\begin{equation}
 \bz_t =  
            \begin{pmatrix}
            0.660   &  0.023 \\
            0.023  &   0.660   
            \end{pmatrix}
            \bz_{t-1} +
            \begin{pmatrix}
            0.109  &  0.015 \\
            0.015 &   0.109   
            \end{pmatrix}
            \bz_{t-2} 
	    \begin{pmatrix}
            0.028  &  0.013 \\
            0.013 &   0.028   
            \end{pmatrix}
            \bz_{t-5} 
            + 
            \begin{pmatrix}
            0.634  &  0.000 \\
            0.013 &   0.634   
            \end{pmatrix}
            \bEpsilon_t 
            .
\end{equation}

\begin{figure}[ht!]
    \centering
    \includegraphics[width=0.85\textwidth]{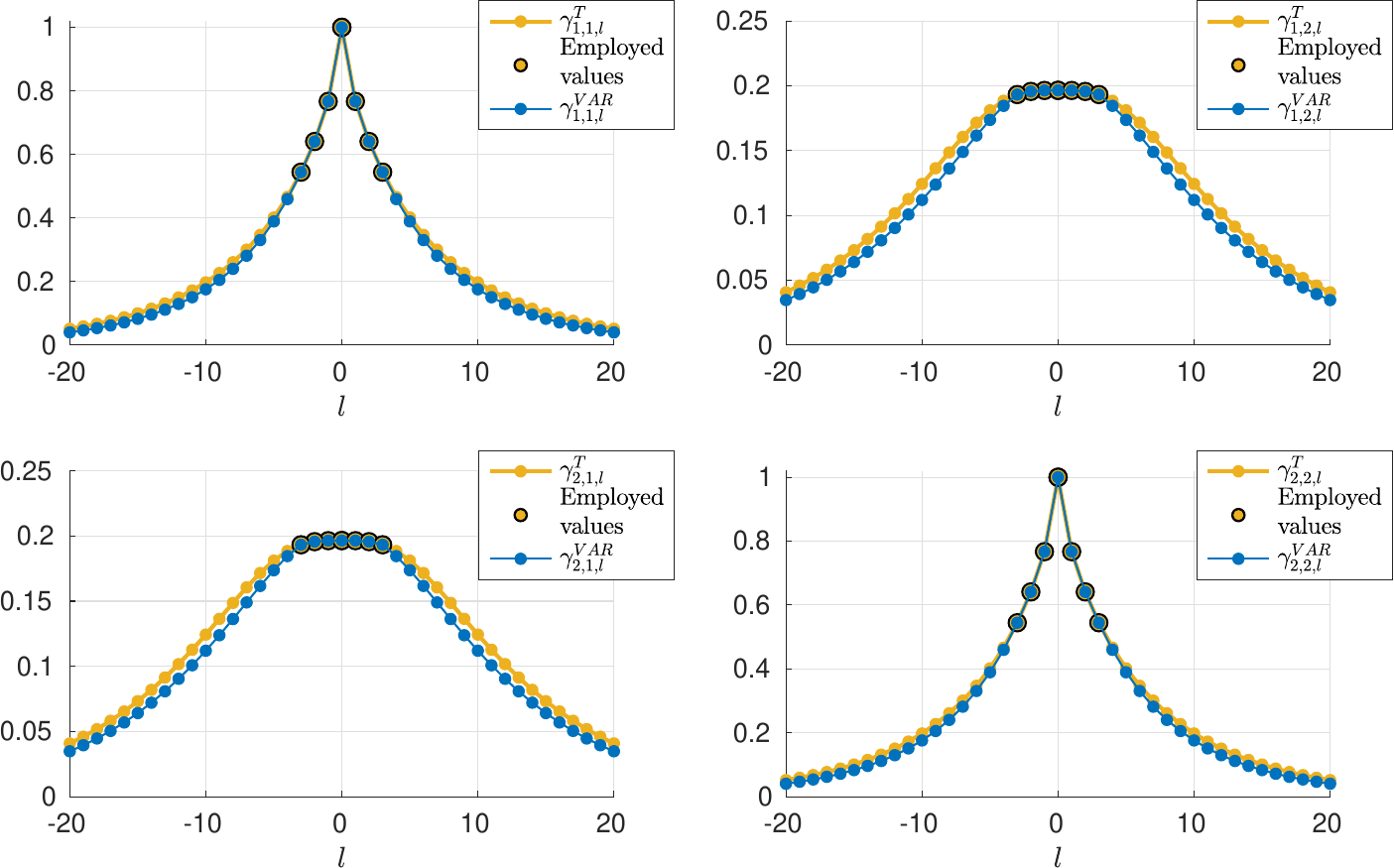}
    \caption{Target and VAR covariance matrix function of a VAR(3) model obtained with the Y-W approach.}
    \label{fig_VAR3_0123}
\end{figure}

\begin{figure}[ht!]
    \centering
    \includegraphics[width=0.85\textwidth]{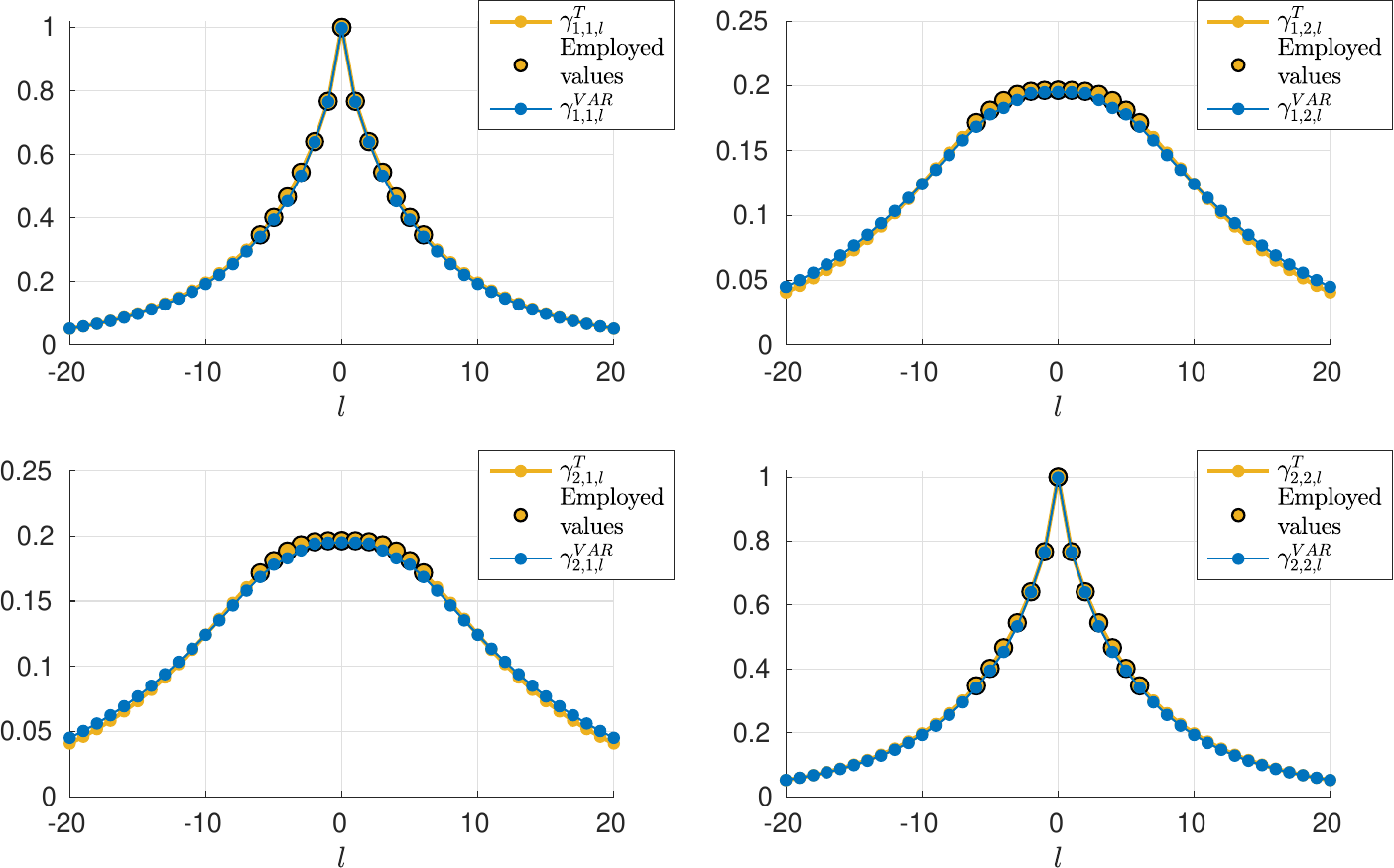}
    \caption{Target and VAR covariance matrix function of a VAR(5,[1,2,5]) model computed for \textbf{l}=[1, 2, 6].}
    \label{fig_VAR125_0126}
\end{figure}

Figures \ref{fig_VAR3_0123} and \ref{fig_VAR125_0126} show the resulting covariance matrix functions, computed through VAR(1) representation (i.e. exact values). As expected, the model obtained with the Y-W approach provides exact matching for lags from $l=-3$ to $l=3$. However, an improved global fitting is obtained with the restricted VAR(5,[1,2,5]) model.

\section{Conclusions} \label{sec_50_conclusions}

Sequential methods for synthetic realisation of random processes have a number of advantages compared to spectral methods. For instance, they are characterised by a more handy synthesis process, as it can be stopped and restarted at any time. In addition, the models obtained through the sequential approach (e.g., autoregressive models) have theoretical expressions of their autocovariance function and power spectrum density (PSD) function, which allows an improved assessment of the accuracy of the models in reproducing the predefined statistical information.

In this article, a methodological proposal for the determination of optimal autoregressive (AR) models from a predefined target autocovariance function was introduced. 
To this end, a general formulation of the problem was developed. Two main features characterise the introduced formulation: (i) flexibility in the choice for the autocovariance equations employed in the model determination, through the definition of a lag vector \textbf{l}; and (ii) flexibility in the definition of the AR model scheme through vector \textbf{j}, that defines the regression terms considered in the model. The AR parameters are directly obtained as a solution of a linear system that depends on \textbf{j} and \textbf{l}. The well-known Yule-Walker (Y-W) and a recent approach based on restricted AR models (K-M) can be seen as particular cases of the introduced formulation, since $\textbf{j}=\textbf{l}=1,...,p$ for Y-W and $\textbf{j}=\textbf{l}$ for K-M.

The introduced formulation was exploited by a genetic algorithm to obtain optimal AR models for the synthetic generation of stationary homogeneous isotropic (SHI) turbulence time series.
The resulting models improved Y-W and K-M based models for the same model parsimony in terms of the global fitting of the target autocovariance function. This achievement was obtained at the expense of reducing the local fitting for some lags. The impact of this trade-off on the frequency domain was presented as a path for extending the notion of optimal model to specific problems in which constraints in the frequency domain may exist, as it is the case in some engineering problems. 

The formulation for the one-dimensional multivariate case was also presented. The reasons behind some computational bottlenecks associated with the multivariate formulation were highlighted. 

Finally, a non-linear approach for the univariate case was described, for which preliminary results suggest an improved fitting of the target autocovariance function.


\appendix

\setcounter{equation}{0}
\renewcommand{\theequation}{\thesection.\arabic{equation}}

\section{Isotropic turbulence} \label{A_isotropic}

The covariance tensor of a three-dimensional statistically stationary, homogeneous and isotropic velocity field between two spatial points separated by vector $\textbf{r}$ is given by:

\begin{equation}
    \textbf{R}(\textbf{r}) = \Cov[ \, \bu(\bx+\br) \, , \, \bu(\bx) \, ] = \E[ \, \bu(\bx+\br) \, \bu(\bx)^T \, ] = \sigma_0^2 \, \left( [f(r)-g(r)] \frac{\br \, \br^T}{\br^T \, \br} + g(r) \, \bI  \right),
\end{equation}

\noindent where $\sigma_0^2$ is the isotropic variance parameter, $r=|\textbf{r}|$, $f(r)$ is the longitudinal correlation function and $g(r)$ is the transverse correlation function \cite{Karman1938}. Assuming incompressibility provides the following relationship between $g(r)$ and $f(r)$:

\begin{equation}
g(r) = f(r) + \frac{r}{2} \frac{\text{d} }{ \text{d} r} f(r)
\end{equation}

For the particular case of a line along the wind direction, $\textbf{r} = [ r , 0 , 0]^T$, the covariance tensor is given by:

\begin{equation}
    \textbf{R}(\textbf{r}) = \begin{pmatrix} 
                    R_{u}(r) & 0 & 0 \\
                    0 & R_{v}(r) & 0 \\
                    0 & 0 & R_{w}(0) \\
                    \end{pmatrix} = ~
					\sigma_0^2 \begin{pmatrix} 
                    f(r) & 0 & 0 \\
                    0 & g(r) & 0 \\
                    0 & 0 & g(r) \\
                    \end{pmatrix} .
\end{equation}

\

From the generalized von Karman spectral density functions, explicit expressions for $f(r)$ and $g(r)$ can be obtained \cite{Krenk2019}:

\begin{equation} \label{eq_f}
    f(r) = \frac{2}{\Gamma(\gamma-1/2)} \, \left( \frac{r}{2L} \right)^{\gamma-1/2} \, K_{\gamma-1/2} \left( \frac{r}{L} \right) ,
\end{equation}

\begin{equation} \label{eq_g}
    g(r) = f(r) - \frac{2}{\Gamma(\gamma-1/2)} \, \left( \frac{r}{2L} \right)^{\gamma+1/2} \, K_{\gamma-3/2} \left( \frac{r}{L} \right) ,
\end{equation}

\noindent where $L$ is the length scale parameter of the three-dimensional von Karman energy spectrum, $\Gamma(\cdot)$ is the gamma function (not to be confused with the covariance function), and $K_n$ is the Bessel function of second kind of order $n$. 
The integral length scale in the longitudinal direction, $L_u^x$, is defined as:

\begin{equation} \label{eq_integral_length_scale}
L_u^x = \int_0^\infty f(r) \text{d} r,
\end{equation}

\noindent and it holds that: 

\begin{equation}
L_u^x = \lambda \cdot L,
\end{equation}

\noindent with $\lambda = \frac{\Gamma(1/2) \, \Gamma(\gamma)}{ \Gamma(\gamma-1/2)}$, where $\gamma$ is a parameter with the value proposed by von Karman, $\gamma=5/6$ (not to be confused with the autocovariance function).

Note that Equation \eqref{eq_integral_length_scale} implies that the integral length scale exists and is $L_u^x < \infty$. However, there is evidence on the fact that geophysical time series, including turbulent velocity fluctuations, may show long-term persistence (Hurst phenomenon), meaning that the integral length scale does not exist \cite{Nordin1972,Helland1978}. According to \cite{Dias2018}, the assumption of a finite integral scale in atmospheric turbulence is not well based on experimental evidence. There are at least two reasons for that: on the one hand, the measurement periods are in the range of minutes-hours due to the inherent non-stationarity of the atmospheric boundary layer. On the other hand, tools for spectrum estimation from records usually include smoothing and low frequency loss of information, which leads to a bad representation of the PSD for very low frequencies (the length scale is the value of the PSD at frequency zero) \cite{Dimitriadis2016}. In \cite{Dimitriadis2016}, a review on the most common three-dimensional power-spectrum-based models of stationary and isotropic turbulence is performed, concluding that Hurst-Kolmogorov (HK) behaviour is systematically excluded. To the author's knowledge, the impact of this fact is still unclear, and it probably depends on the application. For the case of synthetic generation of turbulent wind velocity fields oriented to aeroelastic wind turbine simulation, the impact could be small because the duration of the required simulations is in the order of minutes, and the focus is placed on microturbulence, rather than low frequency fluctuations.

We define the non-dimensional autocovariance function for the longitudinal wind velocity component along the wind direction as a function of the non-dimensional separation $ \mathring{r} = r/L$ as follows:

\begin{equation}
\mathring{R}_u(\mathring{r}) =  \sigma_0^{-2} \, R_u(r=\mathring{r}L)= \frac{2}{\Gamma(\gamma-1/2)} \, \left( \frac{\mathring{r}}{2} \right)^{\gamma-1/2} \, K_{\gamma-1/2} \left( \mathring{r} \right).
\end{equation}

It holds that:

\begin{equation}
\int_0^\infty \mathring{R}_u(\mathring{r}) \, \text{d} \mathring{r}  = \frac{L_u^x}{L} = \lambda.
\end{equation}

The target autocovariance function $\gamma^T_l$ results from taking discrete values from $\mathring{R}_u(\mathring{r}) $ with a given $\Delta \mathring{r}$:

\begin{equation}
 \gamma^T_l =  \mathring{R}_u( l \cdot \Delta \mathring{r})  ~~ , ~~ l=0,1,...
\end{equation}

The numerical application considered in this work was obtained for:

\begin{equation}
 \Delta \mathring{r} = \frac{\lambda}{6} = 0.1245.
\end{equation}

\begin{figure}[ht]
    \centering
    \includegraphics[width=0.75\textwidth]{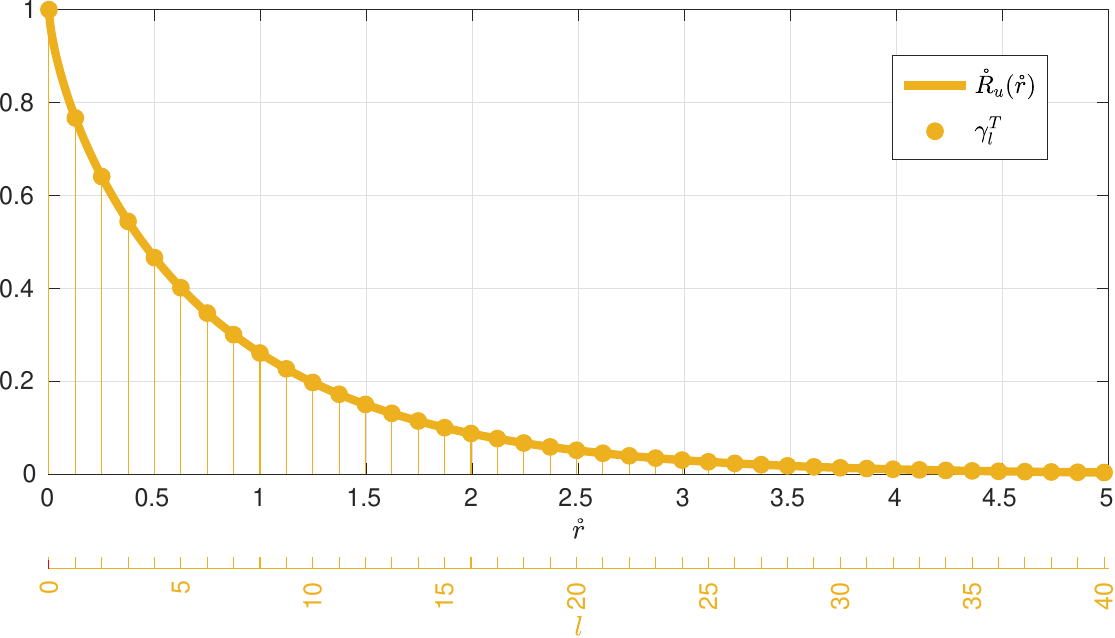}
    \caption{Non-dimensional autocovariance function for longitudinal wind component along wind direction in isotropic turbulence (continuous line), and target autocovariance function for $\Delta \mathring{r} = 0.1245$ (discrete values).}
    \label{fig_Annex_VK}
\end{figure}

Figure \ref{fig_Annex_VK} shows $\mathring{R}_u(\mathring{r}) $ (continuous line) together with $ \gamma^T_l$ (discrete values).

\begin{figure}[ht]
    \centering
    \includegraphics[width=0.75\textwidth]{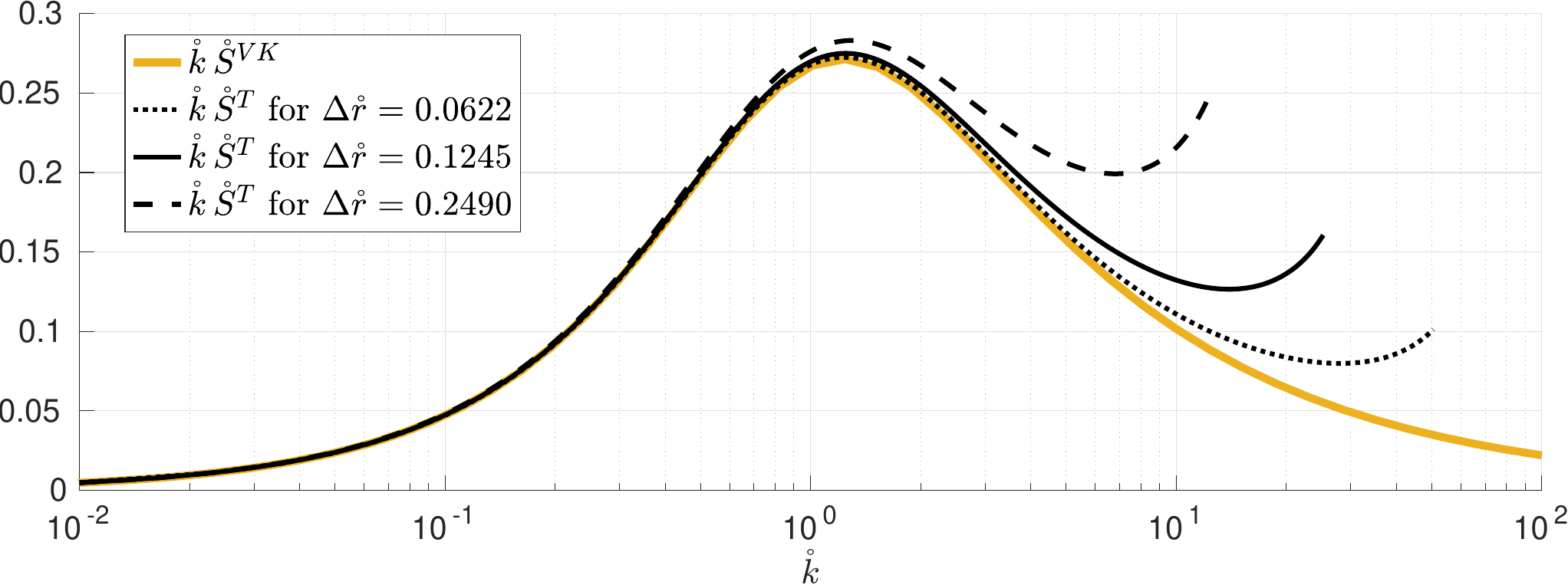}
    \caption{Non-dimensional one-sided von Karman spectrum together with the non-dimensional one-sided target spectrum obtained for different $\Delta \mathring{r}$ values.}
    \label{fig_spectra_and_aliasing}
\end{figure}

It is worth mentioning that the choice for parameter $ \Delta \mathring{r}$ usually represents a trade-off between different criteria \cite{Spanos1983}. On the one hand, small $\Delta \mathring{r}$ values involve target autocovariance functions with non negligible values for a large number of lags, increasing the number of required AR model coefficients to reproduce $\gamma^T_l$ reasonably well. On the other hand, large $\Delta \mathring{r}$ values may derive into problems in reproducing the spectrum for high frequencies (aliasing), which is problematic for some engineering problems \cite{Li1990}. Indeed, by discretising the continuous autocovariance function of a physical process to build $\gamma^T_l$ (see above), the corresponding target spectrum (i.e., the Fourier Transform of $\gamma^T_l$) differs from the spectrum of the original continuous process in the high frequency range. To illustrate this, Figure \ref{fig_spectra_and_aliasing} shows the non-dimensional one-sided spectrum of the employed von Karman turbulence model, $\mathring{S}^{VK}(\mathring{k})$, together with the target spectrum resulting from three different values of $\Delta \mathring{r}$, referred to as $\mathring{S}^{T}$ for a specific $\Delta \mathring{r}$. $\mathring{S}^{VK}(,\mathring{k})$ is given by:

\begin{equation}
\mathring{S}^{VK}(\mathring{k}) =  \frac{S^{VK}(k = \mathring{k}/L)}{\sigma_0^2 \, L}  =  \frac{\Gamma(\gamma)}{\sqrt{\pi}\Gamma(\gamma-0.5)} 
\frac{2 \mathring{k}}{ [1+\mathring{k}^2]^{\gamma} } ~,~ \mathring{k} \in [0,\infty],
\end{equation} 

\noindent where $\mathring{k}$ is the non-dimensional wave number, $\mathring{k}= k \, L$, and $\gamma=5/6$, as stated above. Note that $\mathring{S}^{T}$ are obtained only up to a maximum value of $\mathring{k}$ derived from the Nyquist theorem, $\mathring{k}_{max} = \frac{2\pi}{2 \Delta \mathring{r}}$.

\section{Results for a target autocovariance function with very low decay rate} \label{A_isotropic}

\begin{figure}[ht]
    \centering
    \includegraphics[width=0.75\textwidth]{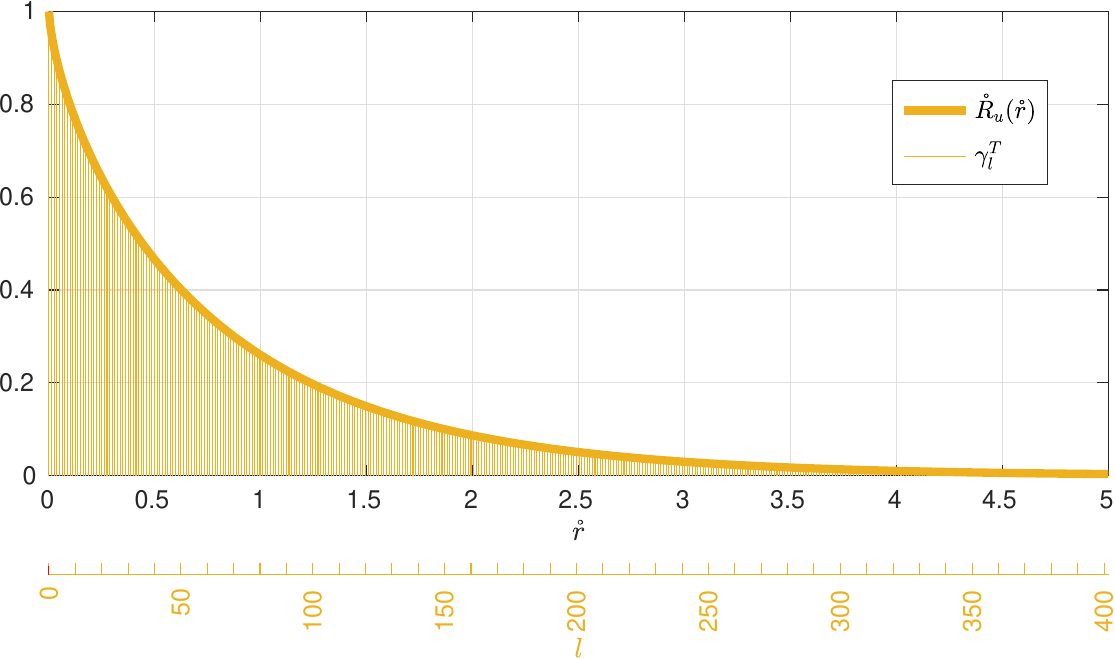}
    \caption{Non-dimensional autocovariance function for longitudinal wind component along wind direction in isotropic turbulence (continuous line), and target autocovariance function for $\Delta \mathring{r} = 0.01245$ (discrete values).}
    \label{fig_Annex_VK_300}
\end{figure}

As mentioned in Section \ref{sec_10_intro}, long-term persistence processes (Hurst phenomenon), characterised by an infinite integral scale, are not considered in this study. However, because many physical processes may show very large integral time scales, it is interesting at least to highlight how the structure of the optimal AR model changes according to the decay rate of the considered target autocovariance function. To this end, the optimisation process performed in Section \ref{sec_40_GAs} has been repeated assuming a discretisation of the non-dimensional continuous autocovariance function ten times finer, i.e. $\Delta \mathring{r} = \frac{\lambda}{60} = 0.01245$. As a consequence, the target autocovariance function decays ten times slower, and the maximum lag considered in the \textit{MSE} computation to account for the $99,5\%$ of the integral length scale becomes $M=401$. Figure \ref{fig_Annex_VK_300} shows the resulting target autocovariance function. A direct consequence of this modification is an increase of the computational burden associated with the optimisation procedure.

Table \ref{table_lenght_scale_increased} shows the obtained AR model order (the last element of vector \textbf{j}) for the different approaches and model parsimony, $N$. While the model order for \textit{Y-W} and \textit{K-M} approaches depends solely on the number of model parameters, $N$, different model orders are obtained for the case of the optimal \textit{GA}-0 and \textit{GA}-10 models, according to the different decay rate of the target.

\begin{table}[h!]
\centering
 \caption{AR model order, $p$, for different approaches, model parsimony, $N$, and decay rates of the target (fast decay: $\Delta \mathring{r} = 0.1245$; slow decay: $\Delta \mathring{r} = 0.01245$)} \label{table_lenght_scale_increased}
 \begin{tabular}{ccccccccc} 
 \hline
 	& 	  &     && \multicolumn{2}{c}{Fast decay of $\gamma_l^T$}  && \multicolumn{2}{c}{Slow decay of $\gamma_l^T$} \\
\cline{5-6}\cline{8-9}
   	& \textit{Y-W} & \textit{K-M} && \textit{GA}-0 & \textit{GA}-10 && \textit{GA}-0 & \textit{GA}-10 \\  
\hline
$N=1$ & 1	 & 1	 && 1	 & 1  && 	1	 & 1  \\
$N=2$ & 2	 & 2	 && 3	 & 3  && 	11	 & 10  \\
$N=3$ & 3	 & 4	 && 5	 & 7  && 	23	 & 42  \\
$N=4$ & 4	 & 8	 && 8	 & 10 && 	37	 & 101  \\
$N=5$ & 5	 & 16	 && 10	 & 12 && 	49	 & 115  \\
$N=6$ & 6	 & 32	 && 13	 & 18 && 	64	 & 132  \\
$N=7$ & 7	 & 64	 && 14	 & 19 && 	77	 & 138  \\
$N=8$ & 8	 & 128	 && 16 	 & 22 && 	86	 & 151  \\
$N=9$ & 9	 & 256	 && 17 	 & 23 && 	94	 & 146  \\
$N=10$ & 10  & 564	 && 19 	 & 23 && 	105	 & 162  \\
 \hline
 \end{tabular}
\end{table}

The comparison of the model performances in terms of the \textit{MSE} is qualitatively similar to that shown in Figure \ref{fig_S_YW_Krenk_GA-0_GA-10}, the single difference being that the error of \textit{K-M} approach stagnates from $N=8$ on, as $\gamma_l^T$ takes values very close to zero at lag $2^8=256$. The optimal models obtained for $N=3$ are shown in Table \ref{table_comparison_models_N_3_300}. A comparison with Table \ref{table_comparison_models_N_3} illustrates the importance of not assuming a predefined AR model structure, as the optimal one clearly depends on the decay rate of the target.

\begin{table}[h!]
\centering
 \caption{AR models obtained for the different approaches for $N=3$ and slow decay rate of the target autocovariance function.} \label{table_comparison_models_N_3_300}
 \begin{tabular}{l l l c} 
 \hline
 Approach   & $~~~\textbf{j}$    & $~~~\textbf{l}$    & Model \\  
 \hline
\textit{Y-W}   & [1,2,3] & [1,2,3]   & $z_t = 0.757 \, z_{t-1} + 0.126 z_{t-2} + 0.080 \, z_{t-3} + 0.310 \, \varepsilon_t $\\
\textit{K-M}   & [1,2,4] & [1,2,4]   & $z_t = 0.757 \, z_{t-1} + 0.138 z_{t-2} + 0.069 \, z_{t-4} + 0.309 \, \varepsilon_t $\\ \textit{GA}-0  & [1,4,23] & [1,4,23] & $z_t = 0.840 \, z_{t-1} + 0.109 z_{t-4} + 0.018 \, z_{t-23} + 0.311 \, \varepsilon_t $\\ \textit{GA}-10 & [1,4,42] & [1,9,34] & $z_t = 0.791 \, z_{t-1} + 0.171 z_{t-4} + 0.009 \, z_{t-42} + 0.310 \, \varepsilon_t $\\ \hline
 \end{tabular}
\end{table}


\section*{Acknowledgements}

This research has been undertaken as a part of the zEPHYR project. This project has received funding from the European Union’s Horizon 2020 research and innovation programme under Grant Agreement No EC grant 860101.

%



%

\section*{Conflict of interest}

The authors declare that they have no conflict of interest.

\bibliographystyle{unsrt}      
\bibliography{bib_tutorial}   

\end{document}